\documentclass[useAMS,usenatbib]{mnras}

\usepackage{amsmath}
\usepackage{amsfonts}
\usepackage{amssymb}
\usepackage{graphicx}
\usepackage[normalem]{ulem}
\usepackage{url}
\usepackage[]{natbib}
\usepackage{RefJ}

\newcommand{\vr}{\vec{r}}

\newcommand{\dd}{\mathrm{d}}

\usepackage{booktabs} 

\usepackage{color}
\definecolor{nicered}{rgb}{.647,.129,.149}

\title[Magnetogenesis Through the Epoch of Reionization]{Mean Energy Density of Photogenerated Magnetic Fields Throughout the Epoch of Reionization}
\author[J.-B. Durrive, H. Tashiro, M. Langer and N. Sugiyama]{Jean-Baptiste Durrive$^{1,2}$\thanks{E-mail:jean.baptiste.durrive@e.mbox.nagoya-u.ac.jp}, Hiroyuki Tashiro$^{2}$, Mathieu Langer$^{1}$ and Naoshi Sugiyama$^{2,3,4}$\\
$^{1}$Institut d'Astrophysique Spatiale, CNRS, UMR 8617, Univ. Paris-Sud, Universit\'e Paris-Saclay, IAS, Bât. 121, 91405, Orsay, France\\
$^2$Department of Physics and Astrophysics, Nagoya University, Nagoya 464-8602, Japan\\
$^3$Kobayashi-Maskawa Institute for the Origin of Particles and the Universe, Nagoya University, Nagoya 464-8602, Japan\\
$^4$Kavli Institute for the Physics and Mathematics of the Universe (Kavli IPMU), The University of Tokyo, Chiba 277-8582, Japan}

\begin{document}

\date{Accepted 2017 August 1; Received 2017 July 15; in original form 2017 May 23}

\pagerange{\pageref{firstpage}--\pageref{lastpage}} \pubyear{2015}

\maketitle

\label{firstpage}

\begin{abstract}
Magnetic fields are ubiquitous in the Universe. They seem to be present at virtually all scales and all epochs. Yet, whether the fields on cosmological scales are of astrophysical or cosmological origin remains an open major problem. Here we focus on an astrophysical mechanism based on the photoionization of the intergalactic medium during the Epoch of Reionization. Building upon previous studies that depicted the physical mechanism around isolated sources of ionization, we present here an analytic model to estimate the level at which this mechanism contributed to the magnetization of the whole Universe, thanks to the distribution of sources, before and alongside early luminous structure formation. This model suggests that the Universe may be globally magnetized to the order of, at least, a few $10^{-20}$~G comoving (i.e.\ several $10^{-18}$~G during the Epoch of Reionization) by this mechanism, prior to any amplification process.
\end{abstract}


\section{Introduction}

Observations indicate that magnetic fields are present in the Universe on a wide range of scales,
from stars through galaxies to galaxy clusters \citep[e.g.][]{Widrow2002,Beck2011,Vallee2011,Feretti2012,Ryu2012,Ferrario2015,Beck2016}.
A relatively recent approach based on the observation of distant blazars repeatedly suggests that intergalactic filaments and cosmic voids too are magnetized to a level that could be at least as high as $10^{-17} - 10^{-15}$ G \citep[e.g.][]{Aleksic2010,Neronov2010,Dolag2011,Tavecchio2011}. Note however that those conclusions depend strongly on assumptions, pertaining to the intrinsic properties of blazars, the extragalactic background light, systematic uncertainties, etc. \citep[e.g.][]{Dermer2011,Arlen2014}. In addition, note that plasma beam instabilities  may be responsible \citep[e.g.][]{Broderick2012,Menzler2015,Chang2016} for the non-detections of secondary electromagnetic cascades that are used to put lower limits on intergalactic magnetic fields \citep[however, for an opposite viewpoint, see][for instance]{Venters2013,Sironi2014,Kempf2016}. On the other side of the strength range, upper limits on intergalactic magnetic fields of the order of $10^{-9}$~G are obtained from the Cosmic Microwave Background,  both temperature and polarization anisotropies and spectral distortions \citep[e.g.][]{Chluba2015,PlanckBFields2016,Zucca2016}, as well as from various large-scale structure observations \citep[e.g.][]{Blasi1999,Pandey2013,Pshirkov2016,Brown2017}.

According to the current paradigm, those fields were first generated as weak seeds
that were later on amplified, perhaps first on small scales within early galaxies through a small-scale dynamo \citep[e.g.][]{2013A&A...560A..87S} or  in the post-recombination intergalactic medium through collisionless plasma instabilities \citep[e.g.][]{FalcetaGoncalves15}. Had they been generated with strengths larger than a few nano-Gauss, magnetic fields would have noticeably affected subsequent structure formation \citep{Wasserman78,Kim96,Tashiro06,Varalakshmi2017}.
Such seed fields were then reorganized through adiabatic
compression and various dynamo mechanisms during or after structure
formation~\citep[e.g.][]{Brandenburg2005,2009A&A...494...21A,Ryu2012}. The origin of the
seed magnetic fields however, particularly on the largest scales, is
still uncertain, despite the many magnetogenesis mechanisms that have
been proposed in the literature \citep[see for instance][for
reviews]{Widrow2002, Kulsrud08, Ryu2012, Widrow12, Durrer2013}. Many of
those mechanisms are based on high energy physics, beyond the standard
model, possibly operating in the early Universe~\citep[see][for recent
reviews]{Widrow12,Durrer2013,Subramanian16}. In the post-recombination
Universe, classical plasma physics is also efficacious to generate
magnetic field seeds through plasma instabilities~\citep[e.g.][]{Gruzinov2001,Schlickeiser2003, Medvedev2006, Lazar2009, Bret2009,Schlickeiser2012}, the \citet{Biermann50} battery \citep[e.g.][]{Pudritz89, Subramanian1994, Ryu1998, Gnedin00,Naoz2013} or the momentum transfer of photons or protons to electrons~\citep[e.g.][]{Harrison1970,Mishustin1972,Harrison1973,Birk2002,Langer03,Langer05,Fenu2011,Saga2015}. Another plausible possibility is that magnetic fields were generated within collapsed structures, and then ejected on larger scales into the intergalactic medium by galactic winds, outflows and AGN jets~\citep[e.g.][]{Rees1987,Daly1990,Kronberg1999,Furlanetto2001,Beck13}.

In~\citet{Durrive15}~(DL15 hereafter; see also~\citealp{Langer05}),
the authors explored in some depth the generation of magnetic fields on
large scales induced by the photoionization of the intergalactic medium
(IGM) during the Epoch of Reionization (EoR). They have shown that,
thanks to ionization-induced charge separation in an inhomogeneous
medium, an electric field possessing a curl component is generated over large distances, thus creating magnetic fields. Typically, $10^{-23}$--$10^{-18}$ G magnetic fields arise on scales which, depending on the nature of the ionizing sources, range from kiloparsecs (Population III stars) and tens of kiloparsecs (primordial galaxies) to megaparsecs (quasars). They have also shown that the scales over which the strengths of the generated magnetic fields are significant are of the order of the average distance between ionizing sources. Thus, this mechanism is naturally able to magnetize the entire Universe at redshifts $z \simeq 30$ to $z \simeq 6$.

In this paper, in order to evaluate the cosmological importance
of this mechanism, we estimate the level of global magnetization of the
Universe it produces. This naturally depends on the distribution of
ionizing sources (namely the typical separation between their
Str\"omgren spheres), their spectral properties, the epochs at which
they appear and the distribution of density inhomogeneities in the IGM. We estimate the mean magnetic strength injected into the IGM by all the sources emitting light above the hydrogen ionization threshold throughout the EoR, with a simple model.

The authors in DL15 modelled the clumpiness of the IGM as a distribution
of baryon overdense clouds. They derived a very detailed formula giving the strength of the magnetic field generated within and around a given overdense cloud surrounding an ionizing source. However the latter expression is rather involved. Fortunately, they also identified the characteristic length scales of the problem
useful for modelling simply the magnetized area around that given
overdensity.
Building upon this, in this paper, we use the
Press-Schechter formalism \citep{Press74} to estimate the statistical
distribution of ionizing sources and of overdensities around these sources.
More precisely, we consider dark matter~(DM) haloes which are massive enough to host luminous
sources and DM overdensity regions which have not yet collapsed but
contain diffuse baryonic overdense clouds.
Then, using an approximate expression for the magnetic field generated
around overdensities in DL15,
we may estimate the magnetic field generated by all the sources forming
during the EoR. Here, for simplicity,
we focus on primordial galaxies only as ionizing photon sources. In fact, it has been suggested that these galaxies are the dominant contributors to reionization \citep[e.g.][and references therein]{McQuinn2016}, a hypothesis recently strengthened by the interpretation of Cosmic Microwave Background data \citep{PlanckCollaboration2016}. Therefore, this approach should give us a realistic estimation of the magnetization level obtained at the end of the EoR.

This paper is  organized  as follows. First, in section \ref{Section:OneSource}, we model the magnetic field generated around one source, due to the presence of one cloud and then due to the presence of a distribution of clouds. Then, in section \ref{Section:DistributionOfSources}, we estimate the global field generated by a distribution of such sources surrounded by clouds during EoR. Conclusions are discussed in section \ref{Section:Discussion}.
Through this paper, we use the Planck reference cosmology~\citep{PlanckCosmoParams2015}, namely $\Omega_b h^2=0.02226$,
$\Omega_c h^2 = 0.1197$ and $h=0.6781$.

\section{Magnetic field generation around one source}
\label{Section:OneSource}

\begin{figure}
\centering
\includegraphics[keepaspectratio, width=1.\linewidth]{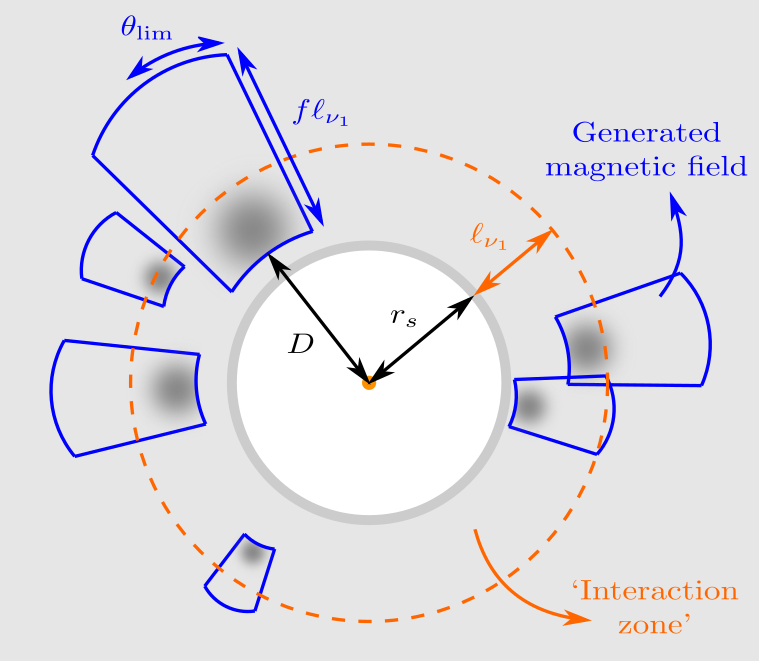}
\caption{Illustration of the magnetic field generated by a luminous
 source~(the orange spot) as it photoionizes the neutral IGM~(the light
 grey region) during EoR. The white region corresponds to the
 Str\"{o}mgren sphere~($r~\leq~r_s$) of the source and the dark grey
 spots are overdense clouds modelling the clumpiness of the IGM. The
 orange dashed line delimits the `interaction zone' ($r_s \leq r \leq
 r_s + \ell_{\nu_1}$), i.e.\ the volume containing the clouds that are
 close enough to the source to participate significantly to the
 magnetogenesis, because further away the number of photoionizations is
 too small. In our calculation we thus take into account only clouds
 inside this zone. Magnetic fields are generated inside and behind these
 clouds, as represented by the blue~frames [corresponding formally to equation~\eqref{Heavisides}], and have strengths well approximated by equation~\eqref{eq:B} (plotted in Fig.~\ref{fig:PlotBApprox}). This modelling abbreviates efficiently the general and detailed results of DL15.}
\label{fig:OneSource}
\end{figure}

\begin{figure}
\centering
\includegraphics[keepaspectratio, width=1.\linewidth]{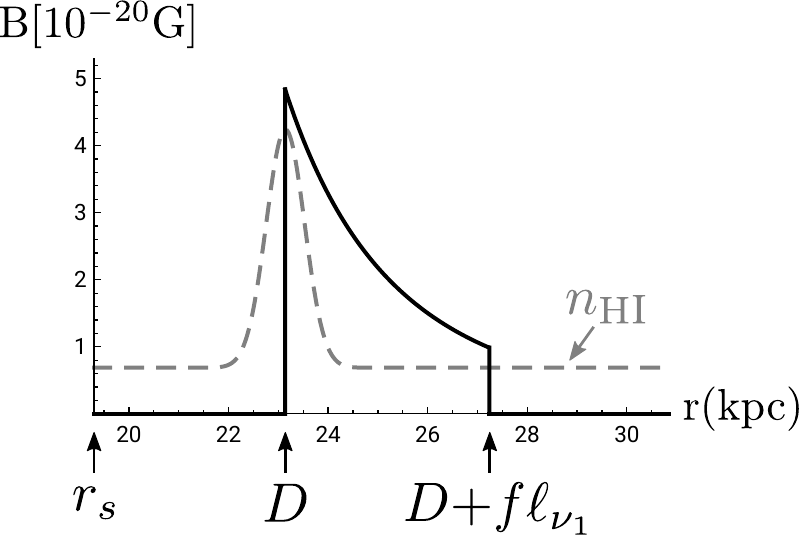}
\caption{Example of the profile given by equation \ref{eq:B}. The grey dashed line is added to illustrate the position and shape of the overdensity (`cloud') inducing this magnetic field.}
\label{fig:PlotBApprox}
\end{figure}

\begin{figure}
\centering
\includegraphics[keepaspectratio, width=1.\linewidth]{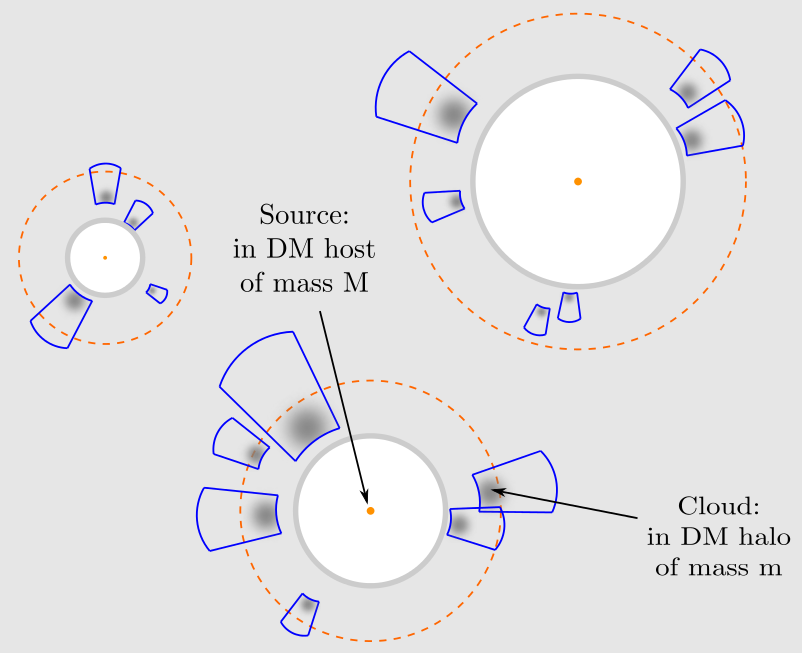}
\caption{This mechanism operates around each source all along the EoR. In this paper we compute the mean magnetic energy density generated (illustrated by the blue frames) in the whole Universe throughout this epoch. We do so by considering that the distribution of sources and the distribution of the (baryonic) clouds surrounding them are given by the distribution of DM haloes in which they are embedded. Conventions in this figure are the same as in Fig.~\ref{fig:OneSource}.
}
\label{fig:DistribSources}
\end{figure}

In DL15 the authors computed in full details (strength, field line
geometries, spatial extent) the magnetic field generated around a single
isolated source as it occasionally ionizes the neutral IGM outside its
Str\"omgren sphere. They showed that three crucial ingredients source
magnetic fields: local inhomogeneities in the electron fraction,
anisotropies of the Str\"omgren sphere and inhomogeneities in the
neutral Hydrogen density. As in this paper, we will focus on the
contribution from inhomogeneities here, leaving the rest for future work. Note that this
already gives us a hint that the estimation deduced here may be an
underestimation of the global field strength (cf.\ section
\ref{Section:Discussion} for more details). Also, in DL15, the authors
performed numerical applications relevant in the cosmological context of
EoR, by considering several types of sources (first stars, primordial
galaxies and quasars), operating at various epochs. One of the outcomes
of their exploration is that one has to be careful with the naive
intuition that, the more powerful the source is, the stronger the
generated magnetic fields are. Indeed, a powerful source emits many photons with short mean free paths (i.e.\ of frequency close to $\nu_0$) generating a large Str\"omgren sphere. Photons with large mean free paths hence first have to propagate a large distance before reaching the neutral IGM (where they participate to the present mechanism). Therefore they are highly diluted (by geometry), which is why for a given spectrum a very powerful source will induce magnetic fields which are in fact weak. Numerically, they
concluded that primordial galaxies constitute the best compromise
between power and dilution. This is interesting since it is precisely
primordial galaxies that are nowadays believed to have been the main drivers
of cosmic reionization. For these reasons, in the present analysis we
will consider only primordial galaxies as ionizing sources.
Following DL15, we assume that sources are characterized by a power-law luminosity in a certain frequency range:
\begin{equation}
L_\nu = L_0 \left(\frac{\nu}{\nu_0}\right)^\alpha \text{ for } \nu \in [\nu_0,\nu_1],
\label{Lnu}
\end{equation}
where $\nu_1$ is the cut-off frequency and we set $\nu_1 = 4 \nu_0$ with $\nu_0$
being the Hydrogen ionization threshold.
Parameters $L_0$ and $\alpha$ depend on the model of ionizing photon
sources. In this paper, the sources considered are primordial galaxies
consisting of Population II stars,
with typically $L_0 = 3 \times 10^{25}$ erg$/\rm s/Hz$ and $\alpha = - 2$. See Sec.~\ref{sec:ionization_IGM} for more details.

\subsection{One cloud}

In DL15, acknowledging that the results they obtained in full generality
were quite involved, the authors have extracted from them the relevant
length scales by considering simple baryon cloud inhomogeneities with a Gaussian profile as toy models. Pursuing in this direction, we will here simplify a little step further their expressions to obtain an efficient but still accurate expression for the fields generated around a cloud near a given source.
As illustrated in Fig.~\ref{fig:OneSource}, we will retain only the
following assumptions: (i) magnetic fields are generated essentially
only by the baryon overdense clouds that are close enough to the source, i.e.\ that are
within what we shall call the `interaction zone', defined as the shell
of thickness $\ell_{\nu_1}$  around the Str\"omgren sphere, where
$\ell_{\nu_1}$ is the mean free path of the most energetic photons
emitted by the source [of frequency $\nu_1$, cf. equation~\eqref{Lnu}],
and (ii) the magnetic field is generated about a cloud at distance $D$ with a simple Gaussian density profile of
neutral hydrogen
\begin{equation}
n_{\mathrm{H\textsc{i}}} = \bar{n} \left(1 + \delta_0 e^{- \frac{(\vr-\vec{D})^2}{2 \sigma^2}}\right),
\label{GaussianInhomogeneity}
\end{equation}
where $\delta_0$ is the central amplitude of the density contrast
(an overdensity for $\delta_0 > 0$, an underdensity for $-1 \leq
\delta_0 < 0$) and $\sigma$ is the width of the cloud,
(iii) the generated magnetic field strength
is well approximated by the following profile~(obtained from equations (39) and (41) of DL15)
\begin{eqnarray}
B_{\sigma,\delta_0,D}(r,\theta,\varphi) &=& B_\mathrm{max}
 \left(\frac{r - r_s + \sqrt{2\pi/e} \ \delta_0 \sigma}{D - r_s +
  \sqrt{2\pi/e} \ \delta_0 \sigma}\right)^{\frac{\alpha-5}{3}}
 \nonumber \\
  && \times \left(\frac{r}{D}\right)^{-3}
 {\cal G}(r, \theta; D,\theta_\mathrm{lim}),
\label{eq:B}
\end{eqnarray}
where $B_\mathrm{max}$ is defined below. This profile is plotted in Fig.~\ref{fig:PlotBApprox}. In the above equation, $\cal G$ is a function delimiting the specific region in which $B$ can be considered as non-negligible. As detailed and illustrated on the right-hand panel of figure $2$ of DL15,
it is relevant to take the following simple expression for the function $\cal G$
\begin{equation}
{\cal G}(r, \theta; D,\theta_\mathrm{lim})=
\Theta(r-D) \Theta(r-r_s) \Theta(\theta_\mathrm{lim} - \theta) \Theta(D+f \ell_{\nu_1}-r),
\label{Heavisides}
\end{equation}
i.e.\ using Heaviside step functions to delimit this region. This is
illustrated by the blue frames surrounding each cloud in
Fig.~\ref{fig:OneSource}.

In DL15, it was shown that the strength of the generated magnetic fields
reaches its maximum at $r=D$. It is given by [cf. equation (39) in DL15]
\begin{equation}
B_\mathrm{max} = t_* \frac{1}{15} \sqrt{\frac{2}{\pi e}} \frac{ \sigma_0^2 L_0 \nu_0}{qx_e D^2} n_\mathrm{H\textsc{I}} \delta_0 F(D, \sigma),
\label{Bdotmax}
\end{equation}
where $t_*$ is the lifetime of the hard photon emitting phase of the source, set to $t_{*} = 100~$Myr in our model, and $F(D, \sigma)$ is the coefficient representing the geometrical effects of the cloud
\begin{equation}
\begin{array}{rl}
 F(D, \sigma) = & \Gamma\left(\frac{5-\alpha}{3}\right) \left(\frac{D - r_s + \sqrt{\pi/2e} \ \delta_0 \sigma}{\ell_{\nu_0}}\right)^{(\alpha-5)/3} \\
& - \Gamma\left(\frac{6-\alpha}{3}\right) \left(\frac{D - r_s + \sqrt{\pi/2e} \ \delta_0 \sigma}{\ell_{\nu_0}}\right)^{(\alpha-6)/3}
\end{array}
\end{equation}
where $\Gamma$ is the gamma function.
For $r<D$ the field is smaller than $B_\mathrm{max}$,
but non-vanishing. However, for simplicity, we take it here equal to zero.
For $r>D$ the strength decays as
the product of power laws in $r$ given by equation~\eqref{eq:B}, but we introduce
a cut-off distance $f \ell_{\nu_1}$, cf. equation~\eqref{Heavisides}, after
which we consider the field to be negligible because it is not physical
to consider infinitely large distances.
Also, the role of the factor $f$ is to let us control this cut-off, measuring it in units of the relevant scale $\ell_{\nu_1}$. Numerically, we observe that the results we obtain below are insensitive to values of $f$ greater than typically~2.

In equation~(\ref{eq:B}),
we are considering that the field is azimuthally symmetric since there is no dependence on the angle $\varphi$, and that the dependence in the angle $\theta$ is piecewise, where the angle $\theta_\mathrm{lim}$ is given by
\begin{equation}
\theta_{\text{lim}} = \arcsin \left(\frac{3 \sqrt{3}}{2} \frac{\sigma}{D}\right).
\end{equation}
Although in reality the magnetic fields have
smooth, angular variations detailed in DL15, these are fair assumptions
since, through the truncation given in equation~\eqref{Heavisides},
 we already consider only a restricted volume where the magnetic
field is generated.

We may now express the energy injected into the IGM in the form of magnetic fields due to one cloud. It is the integral of the magnetic energy density ${B^2}/{8 \pi}$~(Gaussian units) over the whole volume in which the field is generated, that is
\begin{equation}
E_{\sigma,\delta_0}(D) = \int_0^{\theta_\mathrm{lim}} \dd \theta \sin \theta \int_0^{2 \pi} \dd \varphi \int_{D}^{D + f \ell_{\nu_1}} \dd r \ r^2 \frac{B_{\sigma,\delta_0,D}^2}{8 \pi}.
\label{E_OneCloud}
\end{equation}

\subsection{Distribution of clouds}
\label{Section:DistribClouds}

So far we have estimated the magnetic field generated behind a baryon cloud around a
luminous source. However, in reality, an entire distribution of clouds is present around a source,
as illustrated in Fig.~\ref{fig:DistribSources},
and the resultant generated magnetic fields are the sum of the fields
generated by each cloud.
To evaluate the total magnetic fields generated through this mechanism,
it is essential to estimate the distribution of clouds quantitatively.
Therefore, here we make the following assumptions:~(1)
the luminous source is hosted in a DM halo whose mass is noted
$M$ and~(2) each baryon cloud is contained in a DM overdense region
with mass~$m$~which has not yet collapsed.
Hereafter, we call the DM haloes of mass $M$ `hosts'
and the uncollapsed DM overdense regions `DM clouds'.
Note also that, since a DM cloud is only weakly overdense,
we consider that the DM density profile in a DM cloud is the same as that of
the baryons it contains, i.e.\ a Gaussian density profile.

Under this assumption, we can calculate the distribution of the baryon
clouds around a source by considering the distribution of DM overdense
regions around a DM halo.
Based on the definition of the correlation function~\citep{Peebles80},
the probability of finding a DM cloud of mass $m$ within a spherical
shell of volume $4 \pi D^2 \dd D$, at a distance $D$ from a DM host halo of mass $M$, is given by
\begin{equation}
\dd^2 P(D,m|M)  =  \frac{\dd n_m}{\dd m} (1+b_h(M) b_c(m)\zeta(D))\,4 \pi D^2 \dd D \dd m,
\label{eq:Proba}
\end{equation}
where $\dd n_m/\dd m$ is the mass function of the DM clouds, with mass $m$.
To calculate $\dd n_m/\dd m$, we use the Press-Schechter~(PS) formalism~\citep{Press74}.
The function $\zeta$ is the linear matter density correlation function,
and two bias parameters, $b_h(M)$ and $b_c(m)$, are introduced,
respectively, for the host halo of mass $M$ and the DM cloud of mass $m$, to represent the enhancement of these overdensity peaks with respect to the background mass overdensity \citep[cf.][for instance]{MoVdBW}.
To obtain the total magnetic energy generated around the source, we simply need to add the contribution of each cloud. But, as shown in equation~\eqref{E_OneCloud}, the generated magnetic
fields are characterized by the parameters $(\sigma,\delta_0)$ of the profile of the cloud (representing respectively its characteristic size and the value of its central overdensity), and not directly the mass $m$. However, since we
assume that clouds have Gaussian density profiles, we can easily calculate its mass and use parameters~$(m,
\delta_0)$, instead of $(\sigma,\delta_0)$. The PS formalism provides the mass function of clouds with
$(m,\delta_0)$, given the linear density contrast corresponding to $\delta_0$. Now, as mentioned above, while a host is a collapsed object, a cloud is not collapsed. Here we set the critical linear density contrast for a cloud as $\delta_{ch} = 1.05$ corresponding to the turn-around time, and assume that all clouds have the central density
$\delta_0 = 5.55$ corresponding to the non-linear density contrast at the turn-around time.
In the remainder of the paper we will then note $E_m$ the magnetic energy corresponding to $E_{\sigma,\delta_0}$ of equation \eqref{E_OneCloud}. We may then sum up the contribution of all the clouds surrounding the source, and conclude that to each source in a halo of mass $M$ corresponds a magnetic energy
\begin{equation}
E_M = \int_{r_s}^{r_s + \ell_{\nu_1}} \int_{m_\text{min}}^{m_\text{max}} E_m(D)\,\dd^2 P(D, m|M).
\label{eq:E_M}
\end{equation}
The boundaries of the first integral in this equation express
the fact that only the clouds inside the `interaction zone' (cf.\ Fig.~\ref{fig:OneSource}) generate significant magnetic fields and are taken into account. Let us now
discuss the boundaries of the second integral, namely $m_\mathrm{min}$
and $m_\mathrm{max}$ for the mass of the clouds.

For the upper bound, we have two constraints. First, a cloud that is very large may turn out to be totally opaque, even to the most energetic photons emitted by the source. It then does not contribute efficiently to the
magnetization of the IGM since no photon passes through them. More precisely, when light crosses a cloud modelled as a Gaussian overdensity $\delta_0$ of width $\sigma$ embedded in a background density $\bar{n}$, the radiation intensity behind the cloud is attenuated by a factor $\epsilon = \exp \left(- \sigma_{\nu_1} \int n_\mathrm{H\textsc{i}} dr\right)$ with respect to the ambient radiation field. For instance $\epsilon \simeq \exp \left[- (1 + \delta_0) \frac{2\sigma}{\ell_{\nu_0}}\right]$
for $\nu = \nu_0$, and since we fix $\delta_0$, we get a relation between $\sigma$, and therefore on the mass of the cloud, for a given attenuation factor $\epsilon$. In this paper, we set $\epsilon < 1$ as an upper bound for the cloud mass. A second constrain simply comes from the fact that, located at a distance $D$ from the
source of ionizing photons, the width of a cloud cannot be larger than $D-r_s$, otherwise it would encroach on the Str\"omgren sphere of the source. Thus, we take for $m_\mathrm{max}$ in equation~\eqref{eq:E_M} the minimum of these two upper bounds.
On the other hand, very small DM overdensities are unable to host diffuse baryons, due to pressure effects of the latter. Since the sound speed, and thus the Jeans mass may change significantly during the time it takes for an overdensity to grow, \citet{Gnedin1998} showed that the correct mass scale to consider is the so-called `filter mass' $M_\mathrm{F}$ which is of the order of $2 - 3 \times 10^4 \mathrm{M}_\odot$ and varies only a little with redshift until reionization is complete \citep{Naoz2007}. However, the streaming velocity between gas and DM, left over from recombination, may have the effect of increasing $M_\mathrm{F}$ by roughly an order of magnitude \citep[][]{Tseliakhovich2011}. In the following, we take $m_\mathrm{min} = 10^4 \mathrm{M}_\odot$ for simplicity, and we keep in mind that it may be roughly $10$ times larger.
We will discuss the sensitivity of our results to these mass bounds in section \ref{Section:Results}.

\section{Magnetic Energy Density Generated in the IGM}
\label{Section:DistributionOfSources}

In the previous section we have computed explicitly the magnetic energy density generated around an isolated source surrounded by neutral clouds. Now, in order to evaluate the field generated throughout the IGM, we need to take into account the cosmological context in which sources evolve. This consists in three steps.
First, since sources are contained in DM haloes of mass $M$,
we use the PS formalism to estimate their number density.
Second, we need to take into account the fact that
not all DM haloes contain luminous sources.
We introduce in our model the rate at which DM haloes can `switch on' sources,
so as to make it consistent with an important
observational constraint on EoR, namely the optical depth parameter
deduced from the Planck 2015 data \citep{Planck15,PlanckCosmoParams2015}.
Indeed, if too many hosts contain sources, then the EoR ends too soon compared to what observations suggest, and vice versa.
Third, we must account for the fact
that sources switching-on early are isolated, embedded in an essentially neutral medium, and thus generate
the energy computed in the previous section, while those appearing towards
the end of EoR hardly contribute to the magnetization of the IGM because not much of the neutral gas is left
due to the overlapping of cosmological Str\"omgren. We do so by
introducing the ionization fraction of the IGM as follows.

\subsection{Ionization of the IGM}\label{sec:ionization_IGM}

Let us now compute the ionized volume associated with DM haloes.
Assuming a universal baryon-to-dark mass fraction,
a DM halo of mass $M$ contains a mass $M {\Omega_b}/{\Omega_m}  $ of baryons.
However, not all this mass is converted into the stars constituting the ionizing source.
We introduce a parameter $f_*$ representing
the fraction of baryons converted into stars.
With $\dot{N}_\star$, the rate of ionizing photons emitted per baryons,
the ionizing photon production rate from a DM halo with mass $M$~is
given by
\begin{equation}
\dot{N}_{\rm{ion}} = f_* f_{\rm esc} \dot{N}_\star \frac{\Omega_b}{\Omega_m} \frac{M}{m_p},
 \label{eq:dotn}
\end{equation}
where $f_{\rm esc}$ is the escape fraction introduced to account for the fact that
only a fraction of the emitted photons participate to the ionization of the IGM.
The rate $\dot{N}_\star$ depends on the models of ionizing photon sources, but a typical value is $\dot{N}_\star = 40$ Myr${}^{-1}$ (e.g. \citet{LoebFurlanetto13}, which is consistent with the Yggdrasil model\footnote{\url{http://ttt.astro.su.se/~ez/}} used in DL15 \citep{YggdrasilModel}, which uses the \cite{Schaerer2002} and \cite{Raiter2010} single stellar populations).
Once $\dot{N}_{\rm{ion}}$ is obtained for a DM halo of mass $M$,
we get the corresponding luminosity $L_0$ in equation~(\ref{Lnu}) from
\begin{equation}
\dot{N}_{\rm{ion}} = \int_{\nu_0}^\infty \frac{L_\nu}{h \nu} \dd \nu,
\end{equation}
where we set the spectral index $\alpha = -2$.

We evaluate the size of the ionized bubble produced by the source by computing
the corresponding Str\"{o}mgren radius \citep[e.g.][and considerations in DL15]{LoebFurlanetto13}:
\begin{equation}
r_s = \left(\frac{3 \dot{N}_{\rm{ion}}}{4 \pi \alpha_B C n_\mathrm{H\textsc{i}}^2}\right)^{1/3},
\label{eq:strom_r}
\end{equation}
where $\alpha_B$ is the case-B recombination coefficient ($\alpha_B =
2.6\times 10^{-13}~\rm cm^3 s^{-1}$ at a gas temperature of $10^4$ K),
$n_\mathrm{H\textsc{i}}$ is the neutral hydrogen number density in the
IGM, and $C$ is the hydrogen clumping factor. The clumping factor depends on the
redshift and is yet rather poorly constrained. We use the fitting
function $C(z) = 27.466\exp(-0.114z + 0.001328z^2)$ obtained by
\citet{2006MNRAS.372..679M}.
Putting everything together, we can obtain the volume of the ionized bubble produced by a DM halo of
mass $M$ containing ionizing sources, namely
\begin{equation}
V_\text{ion}(M) = \frac{f_* f_{\rm esc} \dot{N}_\star}{\alpha_B C n_{\rm{HI}}^2} \frac{\Omega_b}{\Omega_m} \frac{M}{m_p}.
\end{equation}

We are now ready to compute the ionized fraction of the Universe at a given time
$t$. Ignoring the recombination process inside ionized bubbles,
it is given by the volume filling factor of ionized bubbles,
\begin{equation}
Q_i(t) =  \int^t_{t_0} \dd t \int^{M_\mathrm{max}}_{M_*} \dd M~V_{\rm ion}(M) \, g_{\rm gl} \, \frac{\dd n_M}{\dd M},
\label{Qi}
\end{equation}
where the parameter $g_{\rm gl}$ is the rate at which sources switch on
in DM haloes
and $\dd n_M/\dd M $ is the mass function of DM haloes with mass $M$.
The lower halo mass limit is set by the requirement that the DM hosts contain galaxies.
Thus, the minimum mass in the above equation is the mass of haloes whose virial temperature is below the atomic cooling threshold,
\begin{equation}\label{eq:atomic}
  M_* = 5 \times 10^7 h^{-1}\left(\frac{\mu}{0.6}\right)^{-3/2}\Omega_m^{-1/2}\left(\frac{1+z}{10}\right)^{-3/2}\, \mathrm{M}_\odot,
\end{equation}
where $\mu$ is the mean atomic weight of the gas and $\Omega_m$ the present-day matter density \citep[e.g.][]{Glover2013}.
The time $t_0$ is the time at which the first sources switch on, corresponding to a redshift that we take equal to $z=20$ in the following since we are considering primordial galaxies.

\begin{figure}
\centering
\includegraphics[keepaspectratio, width=1.\linewidth]{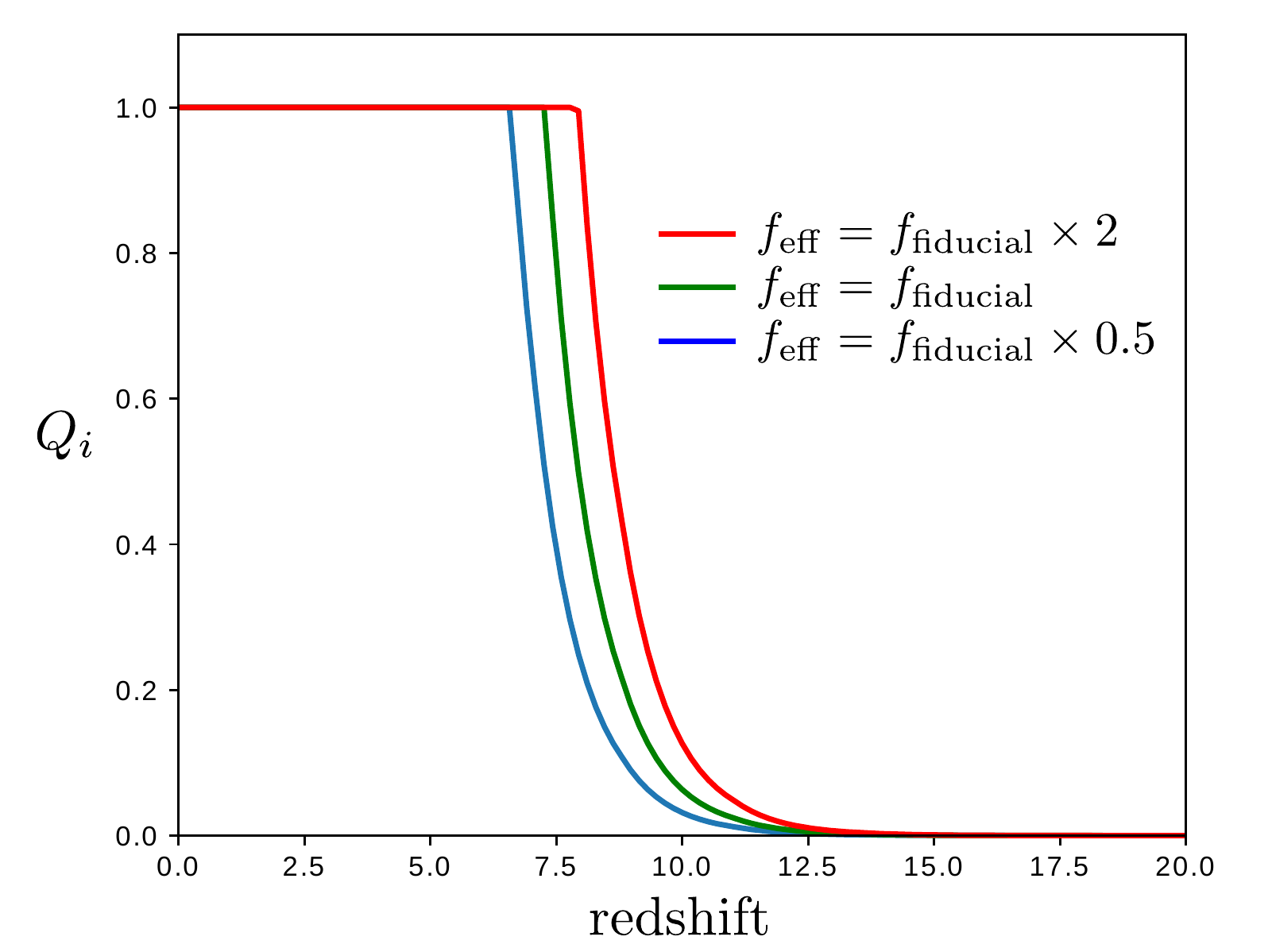}
	\caption{Redshift evolution of the ionization fraction $Q_i$ in our reionization model. The red, green and blue curves are respectively for $f_{\rm eff} =~2\times f_\mathrm{fiducial},~f_\mathrm{fiducial},~0.5\times f_\mathrm{fiducial}$, where $f_\mathrm{fiducial}~=~1.5\times10^{-3}$ is the value of $f_\mathrm{eff}$ in our fiducial model (green curve). The blue curve corresponds to a Universe in which sources are weakly ionizing (low escape fraction and/or low star formation, i.e.\ low $f_{\rm eff}$) while the red curve corresponds to a model with strongly ionizing sources. The blue and red cases are the extreme cases admissible to stay reasonably close to both the measurements of the end of Reionization, and to the measurements of the optical depth (cf. Fig.~\ref{fig:OpticalDepth}).}
\label{fig:IonizFraction}
\end{figure}

\begin{figure}
\centering
\includegraphics[keepaspectratio, width=1.\linewidth]{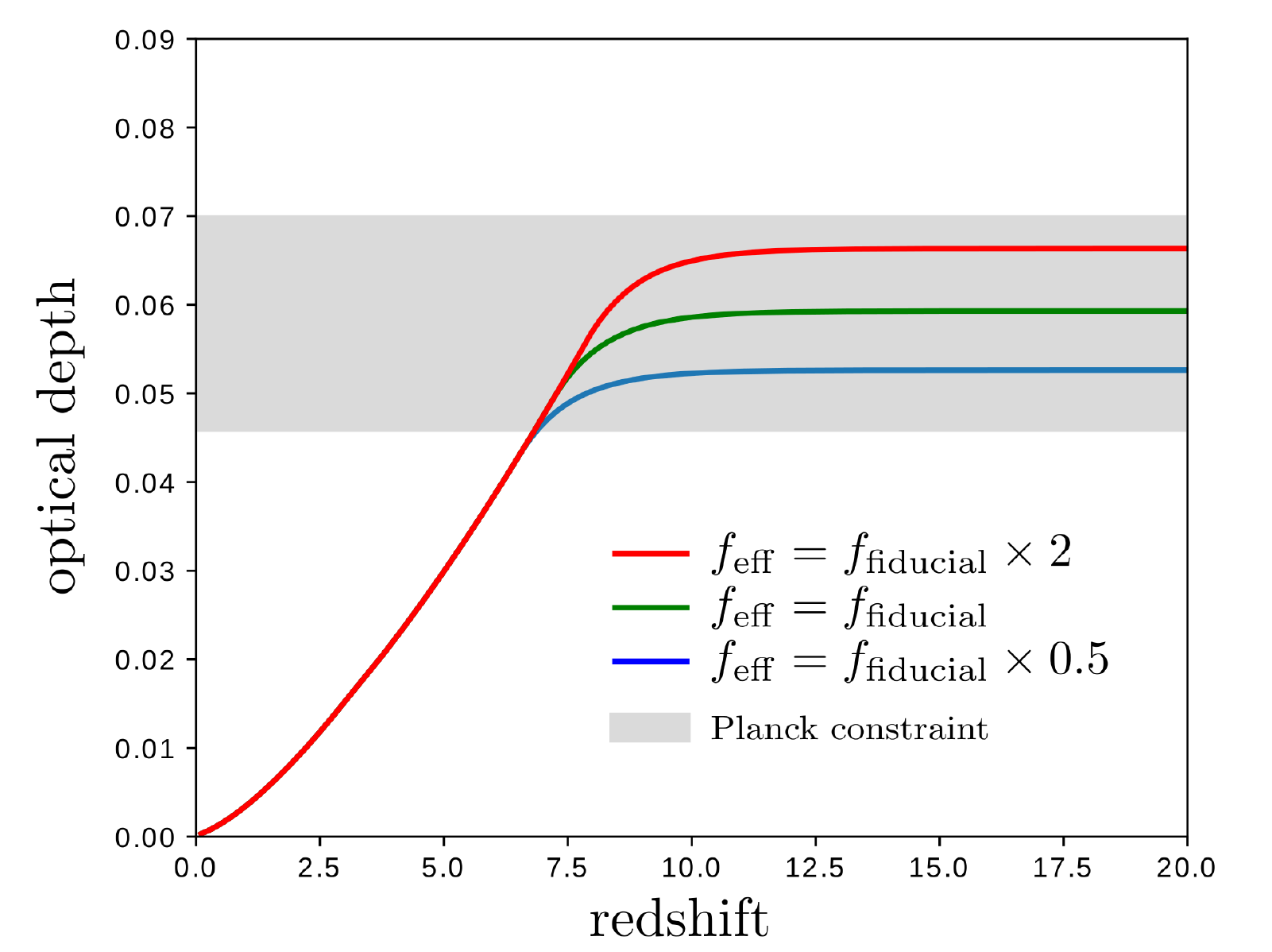}
	\caption{Evolution of the integrated Thomson optical depth $\tau$ to the CMB in our model, in complement to Fig.~\ref{fig:IonizFraction}. We choose values of $f_\mathrm{eff}$ to stay within the error bars of the most recent result $\tau = 0.058 \pm 0.012$ released by the \citet{PlanckCollaboration2016}.}
\label{fig:OpticalDepth}
\end{figure}

To perform numerical applications, we must choose the values of
$f_*$, $f_{\rm esc}$ and $g_{\rm gl}$.
However, the values of the parameters $f_*$ and $f_{\rm esc}$
are uncertain and depend on both the redshift and the source of ionizing
photons. For example, observations of galaxies at $z\sim 3$ by
\citep{Iwata09} indicate an escape fraction of $f_{\rm esc}<0.1$ while
numerical simulations \citep{Wise09,Hayes11,Wise14} suggest that it can be larger than 0.1
 at high redshifts. A natural requirement for these parameters
is that our model of Reionization must be consistent with
the observations and the simulations related to the EoR.
Here, combining these two parameters, we define $f_{\rm
eff} \equiv f_*f_{\rm esc}$ and set $f_{\rm eff} = 1.5 \times 10^{-3}$
in our fiducial model. For the parameter $g_{\rm gl}$, we take it equal to zero at redshifts
greater than $20$, and $g_{\rm gl} = 2 \times 10^{-8}~{\rm yr}^{-1}$ at
$z\leq20$, in order for our fiducial model to be consistent with the
measurements of the ionized fraction during EoR.

Fig.~\ref{fig:IonizFraction} shows the redshift evolution of the
ionized fraction $Q_i$ for different $f_{\rm eff}$.
In the figure, the green line represents
our fiducial model in which the EoR ends at $z=7$.
Note that the value assigned to $g_{\rm gl}$ is in fact quite natural: the corresponding time-scale is $g_{\rm gl}^{-1} = 50$~Myr, and since the period between $z=20$ and $7$ is roughly half a Giga-year long, with this choice we are considering a Reionization driven by about $10$ generations of galaxies.
Of course, taking a redshift independent $g_{\rm gl}$ is a simplification since for instance the metal enrichment process caused by supernovae explosions modifies the galaxy formation rate. However, since we consider galaxies and not Population III clusters, we expect that this metal enrichment process has already been saturated, and therefore it is a fair assumption to take $g_{\rm gl}$ constant with redshift.
For consistency checks, we also computed the Thomson
optical depth to the CMB and, as shown in Fig.~\ref{fig:OpticalDepth}, our fiducial Reionization model is perfectly
consistent with the Planck cosmological result, $\tau = 0.058 \pm 0.012$ \citep{PlanckCollaboration2016}.

In order to explore the role played by the parameter $f_\text{eff}$, we
also consider two other cases.
Red lines in Figs~\ref{fig:IonizFraction} and \ref{fig:OpticalDepth}
correspond to a Universe in which galaxies are
strongly ionizing, i.e.\ they emit ionizing photons at high rates (high
$\dot{N}_\text{ion}$, high $f_\text{eff}$), either because stars are
formed very efficiently (high $f_*$) or because photons are not trapped
(high $f_\text{esc}$).
It is thus natural to see in Fig.~\ref{fig:IonizFraction} that they
reionize the Universe faster than in the fiducial model, and that in
this case the optical depth is larger since more electrons are freed
sooner. On the other hand, blue lines correspond to the opposite situation.
Fig.~\ref{fig:OpticalDepth} shows that, in both cases, the optical depth
remains within the error bars of the Planck recent results.

\begin{figure}
\centering
\includegraphics[keepaspectratio, width=1.\linewidth]{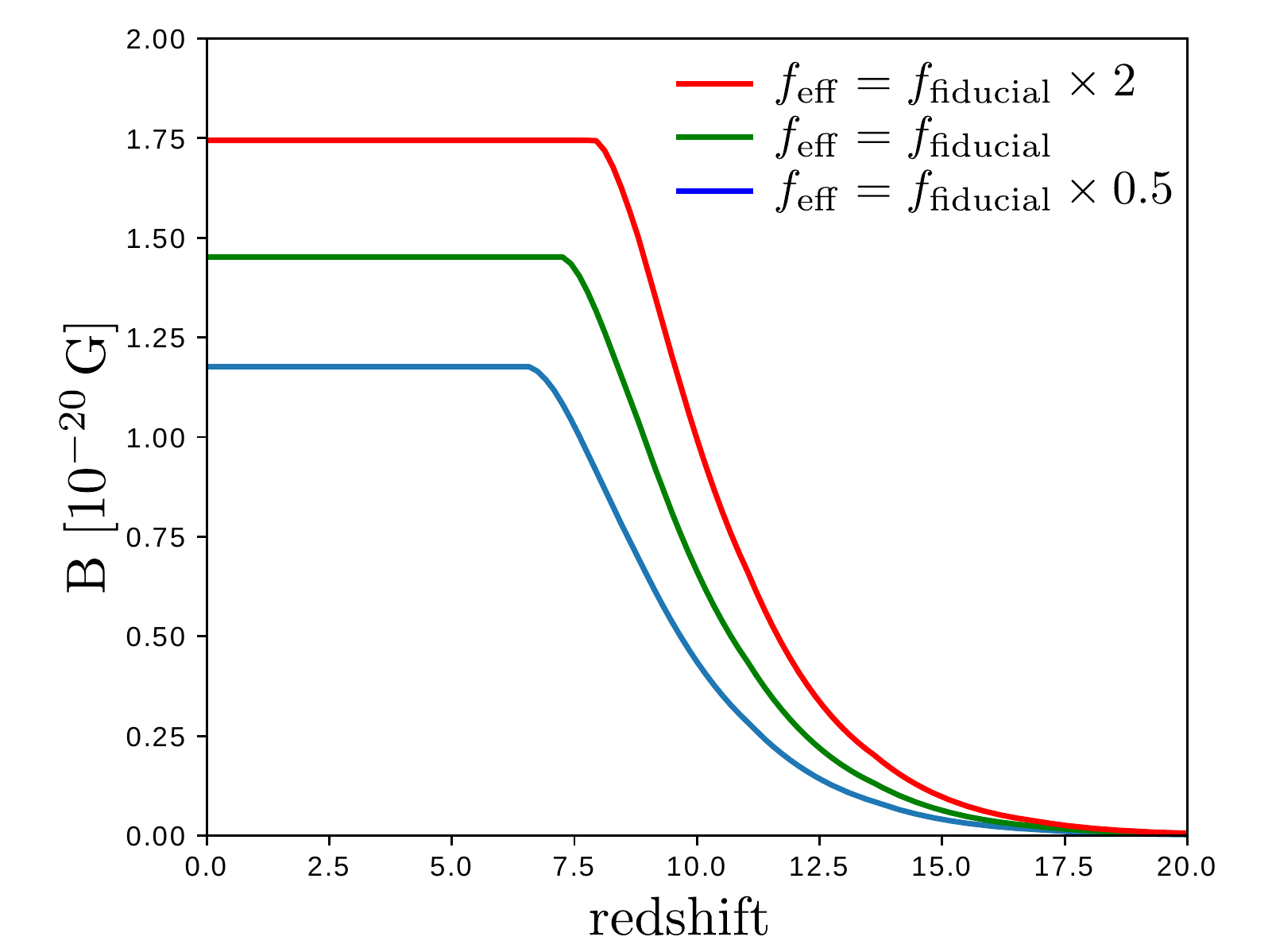}
	\caption{Evolution with redshift of the mean comoving magnetic field accumulated in the IGM generated by the first galaxies for different reionization parameters, computed from formula \eqref{eq:b2pz}. As detailed in Figs~\ref{fig:IonizFraction} and \ref{fig:OpticalDepth}, the green curve corresponds to our fiducial model, the blue curve corresponds to a Universe in which sources are weakly ionizing, and the red curve to strongly ionizing sources. Note that these curves correspond to comoving values, so that the physical magnetic strength is of several $10^{-18}$~G during EoR, without taking any amplification process into account.
	}
\label{fig:PlotB}
\end{figure}

\subsection{Results}
\label{Section:Results}

We can now calculate the mean magnetic energy density (physical) generated by photoionizations
during the EoR.
For simplicity, we assume that
magnetic flux is frozen in the IGM and we do not
consider any possible dissipation or amplification.
Thus, magnetic fields are continuously generated in the IGM
during the EoR and are subsequently diluted adiabatically by cosmic expansion once they reach their maximum strength.

Accordingly, with equations \eqref{eq:dE}, \eqref{eq:dtoverdz} and \eqref{eq:Source},
we eventually obtain that the mean physical magnetic energy density generated
by photoionizations during EoR evolves with redshift as
\begin{equation}\label{eq:b2pz}
    \frac{B_\mathrm{p}^2(z)}{8 \pi} = (1+z)^4 \int^{z_0}_z \dd z' \frac{1-Q_i}{(1+z')^5H}\int^{M_\mathrm{max}}_{M_*} \dd M E_M g_\mathrm{gl}~\frac{\dd n_M}{\dd M},
\end{equation}
where $z_0$ (taken equal to $20$ here) is the redshift at which the entire process sets in (full derivation provided in section \ref{B-evol}).
Since by assumption there are no magnetic fields initially, we have set $B_\mathrm{p}(z_0) = 0$.

Let us explain where the $1-Q_i$ factor comes from. The magnetic energy calculated in equation \eqref{eq:E_M}, giving the energy in magnetic fields
generated around a single source, must be integrated over the distribution of DM haloes.
However, ionized bubbles of individual sources start to overlap around the end of the EoR.
Since our magnetic field generation mechanism takes place only in the neutral IGM, the efficiency of magnetic field generation actually decreases as the
reionization process proceeds. Accordingly, the factor $1-Q_i(z)$ reduces the generated magnetic field energy  as time increases, and terminates the whole process when Reionization is completed.

In Fig.~\ref{fig:PlotB}, we plot
the comoving strength of the generated magnetic field, $B_\mathrm{c} = B_\mathrm{p}/(1+z)^2$, as a function of redshift, with various Reionization histories (the various curves corresponding to those in Figs~\ref{fig:IonizFraction} and \ref{fig:OpticalDepth}).
The global trend of each curve can be naturally explained as follows. Above $z=20$ there are no galaxies, so that the strength is equal to zero, and then as time passes, galaxies form and their radiation generates magnetic fields, i.e.\ a tiny fraction of their radiation energy is converted into magnetic energy, which accumulates in the IGM, so that the curves rise with decreasing redshift. Once the Universe is totally ionized, the generation stops and a plateau is reached.

An interesting feature appears when comparing
Figs~\ref{fig:IonizFraction} and~\ref{fig:PlotB}.
For redshifts above roughly $10$, the Universe is still mostly neutral ($Q_i \ll 1$), and yet for $z$ between $15$ and $10$ a significant fraction of the final total magnetic field strength is already generated. It means that the first sources are quite efficient at generating magnetic fields in the IGM. This is consistent with the fact that this mechanism operates in neutral regions: at these redshifts, the Str\"omgren spheres do not overlap yet, and the mechanism around each source works maximally. On the contrary, later on, less magnetic field is generated relatively to the number of sources which ionize the IGM.

Comparing the three curves in Fig.~\ref{fig:PlotB},
we observe the following behaviour:
the more strongly ionizing the sources are
[corresponding to higher $f_\text{eff}$, i.e.\ either stars form very
efficiently (high $f_*$) or photons are not trapped (high
$f_\text{esc}$)], the more magnetic fields are generated.
Although it may seem obvious, it is not so simple because of the two
following reasons.
First, the stronger the sources, the more efficiently they
reionize the Universe and the shorter the duration of the EoR is
 (and indeed, the plateau is reached earliest in the red case,
and latest in the blue case). There is thus less time for the mechanism to
operate. Secondly, as mentioned in the beginning of section \ref{Section:OneSource}, one
should be cautious with the `naive' and incorrect intuition that more
powerful sources should generate stronger fields, since what matters
in this mechanism is the compromise between having a high rate of
photoionizations but also a high density of photons.
In fact, the behaviour mentioned above can be accounted for as follows.
In the case of the red curve, because the Universe is reionized early
(cf. Fig.~\ref{fig:IonizFraction}) the sources form in a medium with
high density, and thus have small Str\"omgren spheres, so that in this
case they generate strong magnetic fields.
Since the neutral hydrogen density decreases as the Universe expands,
the magnetic field generations are less efficient in the green and blue
cases in which the source production is delayed with respect to the red case.

At the end of this section, it is worth mentioning how the total strength of the field (i.e.\ the level of the plateau
in Fig.~\ref{fig:PlotB}) depends on other parameters in our models.
The most crucial ones are the boundaries of the integrals containing the
mass parameters $m$ and $M$, since they determine which clouds and which
sources can contribute to the magnetic field generation,
through  equations \eqref{eq:E_M}, \eqref{Qi} and \eqref{eq:b2pz}.
At the high end of the mass range, it turns out that the
precise value of $M_\mathrm{max}$ does not matter as long as it is large enough (in our
model, we take it equal to $10^{16} M_\odot$) since the mass function
plummets at such large masses. Conversely, as previously mentioned, the lower boundary $M_*$ for $M$ is directly fixed by the physics of atomic cooling for star formation to be effective.
For the mass $m$ of the DM clouds containing neutral gas, the constraints on the upper bound discussed in section \ref{Section:DistribClouds} are not very constraining. Indeed, if in fact one takes into account clouds with exaggerated (non-physical) sizes, the result is not very much changed because these overdensities are so large that they have very small density gradients, which is the key element of the present magnetogenesis mechanism, and thus do not generate important magnetic fields.
For the lower limit of the parameter $m$, while we already provided physical arguments to constrain its value in section
\ref{Section:DistribClouds} to $10^4\, \mathrm{M}_\odot$, we also evaluated its impact on the
total strength of the magnetic field by exploring a reasonably broader range of values.
We found that the total magnetic field strength
gradually increases as $m_\mathrm{min}$ goes down.
This is natural since in that case we are  taking more and smaller neutral gas clouds into account.
Decreasing $m_\mathrm{min}$ down to the (redshift dependent) Jeans mass increases the resulting magnetic strength by less than 16 percent with respect to the fiducial value. On the contrary, including the effect of the streaming velocity between baryons and DM increases the filtering mass roughly by an order of magnitude. Since this suppresses neutral gas clouds in small DM haloes, it decreases the resulting magnetic strength, roughly by a factor of $2$.

\section{Discussion}
\label{Section:Discussion}

Clearly, our model for the global magnetization of the Universe during the EoR is simplistic in several aspects and may be improved in different ways. However, given the high complexity of the underlying physics and context, the advantage of this analytic approach is that it helps disentangling the problem by explicating the main elements that determine the overall numerical value of the strength of the field, and provides us with an understanding of how they are at play, before resorting to numerical simulations. It is also an important first approach to the problem because it offers the valuable advantage of showing where the difficulties in the modelling are, in the perspective a more refined approach.

Let us now briefly discuss points that we have not directly addressed in our
model and try to assess whether they result in overestimating or underestimating the generated field.
Note however that despite these, this work already gives a pertinent hint of the correct order of magnitude.

It has been shown in DL15 that the physical mechanism at the heart of our model is more efficient in underdense, rather than overdense regions of the IGM. Indeed, photons are less absorbed there, so they travel further and interact within more extended regions. Also, at a given distance from the source and with a given density gradient, the strength of the locally generated field  is larger in the underdense case than in the overdense case since more photons reach that distance. In this paper, for simplicity, we considered only overdensities to model the clumpiness of the IGM. We therefore neglected the a priori important contribution of the underdense regions between those clumps. A precise modelling of the profile of those regions is out of the scope of this work, but crudely speaking, since in our present model we did not consider roughly half of the neutral density gradients in the IGM, we estimate that the values of the generated field derived here could be actually doubled due to this.

In DL15, the authors justify that since the mechanism is assumed to operate around single sources during $100$ Myr, it is relevant to consider, in a first approach, that the Str\"omgren spheres have reached their steady state. We followed this approach in this paper. However, whether this results in an overestimation or an underestimation of the overall mean generated magnetic field is not obvious. Indeed, in the transitory phase during which the Str\"omgren sphere grows, photons reaching the IGM are less geometrically diluted, so that the mechanism is very efficient closer to the source. At the same time, it is then less efficient further away, since photons above the hydrogen ionization threshold are in this case more absorbed than once the sphere has reached its steady state size. This work implicitly assumes that these two effects average out when we estimate the mean magnetic field, sticking to statistical information only.

In equation~\eqref{eq:b2pz}, we introduced a factor $1-Q_i$ to take into account the fact that the fraction of neutral gas decreases as sources switch on. However, we used equation~\eqref{eq:E_M} for the generated energy around each source, which is in principle only valid for an isolated source. To refine further the model, we would need to take into account the effect of neighbouring sources. This would not be as straightforward as it may seem, since we would need to model carefully how the field is generated in and around clouds that are illuminated by multiple sources, i.e.\ for instance how equation~\eqref{eq:B} for the generated field is modified when the radiation field is not unidirectional. This is best investigated probably through numerical simulations, which are actually under study for a forthcoming publication~\citep[][in preparation]{Durrive17}.

In this paper, we have not considered all the elements sourcing
magnetic fields by photoionization. Indeed, as detailed in DL15, local
inhomogeneities in the electron fraction as well as asphericity
of the Str\"omgren regions also contribute to generating magnetic fields in the neutral IGM during EoR, in addition to the neutral hydrogen inhomogeneities that we have considered here. Asphericities of the Str\"omgren regions are probably a very important contribution since they induce potentially strong differences between adjacent lines of sight\footnote{Such differences are  key for the electric field to possess a curl, thus for the induction of magnetic fields, as detailed in DL15.} directly near the ionization front, i.e.\ where the photon number density in the IGM is the largest. Once these transverse gradients exist, they are, in general, maintained all along the radial direction from the source. Therefore, we expect that contribution to yield non-negligible magnetic strengths not only locally, but also on extended distances. This is another reason to hypothesize that the value of the global magnetic field derived in this paper may be quite an underestimate.

In comparison with other astrophysical processes operating during the EoR, let us first mention the work of \citet{Subramanian1994} and \citet{Gnedin00}.
These authors explored the efficiency of the Biermann battery operating as ionization fronts travel through
overdensities. The values we obtain for the generated magnetic field are
just slightly lower than the values obtained in the \citet{Gnedin00}
study, and somewhat larger than those of
\citet{Subramanian1994}. However, in those studies, fields with
significant strengths are generated essentially locally, in dense
structures while, as already emphasized in DL15, the present mechanism,
based on momentum transfer between photons and initially bound
electrons, naturally generates the magnetic seeds deep in the neutral
IGM, into which X and UV photons actually penetrate. Also,
in~\citet{Gnedin00}, the strengths  reported in protogalactic structures
are reached after amplification due to gas compression has taken
place. The values we have obtained here are obtained directly, \textit{without additional processing}.
Though turbulence in the neutral IGM during the EoR is not well observationally constrained yet, simulations have shown that copious amounts of turbulent motions arise naturally during the formation of the first galaxies \citep[][]{Greif2008,Sur2012}. Similarly, mechanical feedback associated with the formation and evolution of large-scale structure in the post-reionization universe does inject turbulence into the IGM \citep[][]{Rauch2001,Ryu2008,Oppenheimer2009,Evoli2011,Iapichino2011,Ravi2016}. Stretching, twisting and folding of magnetic field lines associated with the compressive and shearing motions of turbulence will in reality amplify and reorganize the seed fields obtained in this study, and bring them to the strengths detected in the present-day, structured IGM \citep[e.g.][and references therein]{Ryu2008,Schleicher2010,Sur2012,Vazza2014}. Within cosmic voids, collision-less plasma instabilities have the potential to amplify rapidly the magnetic seed fields \citep[e.g.][]{FalcetaGoncalves15}, and bring them to and maintain them above the lower limits suggested by the observation of distant blazars and cited in the Introduction section \citep[see also][]{2015ApJ...814...20F}.

Another interesting process that may begin already during the EoR is the spontaneous emission of aperiodic turbulent magnetic field fluctuations in the initially unmagnetized intergalactic plasma \citep{Schlickeiser2012a,Schlickeiser2012,2013ApJ...778...39S}. Taking into account viscous damping from collisional processes, those fluctuations may reach strengths of the order of $10^{-12}$ G in protogalactic clouds, where they can contribute to seeding the amplifying dynamo actions relevant to galactic magnetic fields. In the post-reionization plasma of cosmic voids they may reach $10^{-21}$ G \citep{Schlickeiser2012a}. Those strengths can be larger, up to the $10^{-16}$ G level in fully ionized regions (in partially ionized regions, a somewhat stronger damping reduces that level), when collective effects enter the game \citep{2013ApJ...778...39S}. However, while this mechanism may operate throughout the entire IGM, the resulting magnetic field fluctuations are generated on very small scales, smaller than $10^{-4}$ pc, on time-scales of the order of $10^{10}$ years. On the contrary, the mechanism of global IGM magnetization examined in this paper relies on the detailed process of photon-to-electron momentum transfer that creates magnetic field seeds on scales comparable to the distance between ionizing sources, as detailed in section \ref{Section:DistributionOfSources}, and with a coherence essentially set by the size of the gas inhomogeneities present in the IGM \citep[see also][]{Durrive15}. In addition, it magnetizes the entire IGM, including those regions that will become cosmic voids, by the end of the EoR, that is within the first billion years of the Universe.

In conclusion, our model suggests that the Universe may be globally magnetized to the order of \textit{at least} a few $10^{-20}$ G (comoving) by this mechanism\footnote{Potentially larger values could be reached due to possible kinetic effects \citep[][]{Munirov2017}.}. Note that this order of magnitude falls within the range of values obtained in the numerical applications performed in DL15 for clouds close to a single source. This work thus shows that the strength of the fields generated by this mechanism is not only important locally, i.e.\ around isolated sources, but also in a global context, i.e.\ that the typical distribution of sources, the clumpiness of the IGM and the typical duration of EoR, allow for this mechanism to be of cosmological relevance. As a final note, it is an exciting perspective that such magnetic seeds might be directly measurable with the Square Kilometre Array by means of the method based on 21-cm tomography proposed in \citet{Venumadhav2017} and \citet{Gluscevic2017}.

\section*{Acknowledgments}

The authors are grateful to the organizers of the conference and workshops `Origin, Evolution, and Signatures of Cosmological Magnetic Fields' held in June-July 2015 at NORDITA where this work was initiated.
JBD acknowledges financial support by the P2IO LabEx (ANR-10-LABX-0038) in the framework `Investissements d'Avenir' (ANR-11-IDEX-0003-01) managed by the French National Research Agency (ANR). This work was supported in part by JSPS Grants-in-Aid for Scientific  Research under Grant Nos. 25287057 (N.S.), 15H05890 (N.S.) and 15K17646  (H.T.).

\bibliographystyle{mnras}
\bibliography{MTEoR}

\begin{thebibliography}{}
\makeatletter
\relax
\def\mn@urlcharsother{\let\do\@makeother \do\$\do\&\do\#\do\^\do\_\do\%\do\~}
\def\mn@doi{\begingroup\mn@urlcharsother \@ifnextchar [ {\mn@doi@}
  {\mn@doi@[]}}
\def\mn@doi@[#1]#2{\def\@tempa{#1}\ifx\@tempa\@empty \href
  {http://dx.doi.org/#2} {doi:#2}\else \href {http://dx.doi.org/#2} {#1}\fi
  \endgroup}
\def\mn@eprint#1#2{\mn@eprint@#1:#2::\@nil}
\def\mn@eprint@arXiv#1{\href {http://arxiv.org/abs/#1} {{\tt arXiv:#1}}}
\def\mn@eprint@dblp#1{\href {http://dblp.uni-trier.de/rec/bibtex/#1.xml}
  {dblp:#1}}
\def\mn@eprint@#1:#2:#3:#4\@nil{\def\@tempa {#1}\def\@tempb {#2}\def\@tempc
  {#3}\ifx \@tempc \@empty \let \@tempc \@tempb \let \@tempb \@tempa \fi \ifx
  \@tempb \@empty \def\@tempb {arXiv}\fi \@ifundefined
  {mn@eprint@\@tempb}{\@tempb:\@tempc}{\expandafter \expandafter \csname
  mn@eprint@\@tempb\endcsname \expandafter{\@tempc}}}

\bibitem[\protect\citeauthoryear{Aleksi{\'{c}} et~al.,}{Aleksi{\'{c}}
  et~al.}{2010}]{Aleksic2010}
Aleksi{\'{c}} J.,  et~al., 2010, \mn@doi [Astronomy {\&} Astrophysics]
  {10.1051/0004-6361/201014747}, 524, A77

\bibitem[\protect\citeauthoryear{Arlen, Vassilev, Weisgarber, Wakely  \&
  Shafi}{Arlen et~al.}{2014}]{Arlen2014}
Arlen T.~C.,  Vassilev V.~V.,  Weisgarber T.,  Wakely S.~P.,   Shafi S.~Y.,
  2014, \mn@doi [The Astrophysical Journal] {10.1088/0004-637X/796/1/18}, 796,
  18

\bibitem[\protect\citeauthoryear{{Arshakian}, {Beck}, {Krause}  \&
  {Sokoloff}}{{Arshakian} et~al.}{2009}]{2009A&A...494...21A}
{Arshakian} T.~G.,  {Beck} R.,  {Krause} M.,   {Sokoloff} D.,  2009, \mn@doi
  [\aap] {10.1051/0004-6361:200810964}, \href
  {http://adsabs.harvard.edu/abs/2009A%26A...494...21A} {494, 21}

\bibitem[\protect\citeauthoryear{Beck}{Beck}{2011}]{Beck2011}
Beck R.,  2011, in Aharonian F.~A.,  Hofmann W.,   Rieger F.~M.,  eds, 25th
  Texas Symposium on Relativistic Astrophysics (Texas 2010). AIP Conference
  Proceedings.
pp 117--136 (\mn@eprint {} {1104.3749}), \mn@doi{10.1063/1.3635828}

\bibitem[\protect\citeauthoryear{Beck}{Beck}{2016}]{Beck2016}
Beck R.,  2016, \mn@doi [The Astronomy and Astrophysics Review]
  {10.1007/s00159-015-0084-4}, 24, 4

\bibitem[\protect\citeauthoryear{{Beck}, {Hanasz}, {Lesch}, {Remus}  \&
  {Stasyszyn}}{{Beck} et~al.}{2013}]{Beck13}
{Beck} A.~M.,  {Hanasz} M.,  {Lesch} H.,  {Remus} R.-S.,   {Stasyszyn} F.~A.,
  2013, \mn@doi [\mnras] {10.1093/mnrasl/sls026}, \href
  {http://adsabs.harvard.edu/abs/2013MNRAS.429L..60B} {429, L60}

\bibitem[\protect\citeauthoryear{{Biermann}}{{Biermann}}{1950}]{Biermann50}
{Biermann} L.,  1950, Zeitschrift f\"ur Naturforschung A, \href
  {http://cdsads.u-strasbg.fr/abs/1950ZNatA...5...65B} {5, 65}

\bibitem[\protect\citeauthoryear{Birk, Wiechen  \& Lesch}{Birk
  et~al.}{2002}]{Birk2002}
Birk G.~T.,  Wiechen H.,   Lesch H.,  2002, \mn@doi [Astronomy and
  Astrophysics] {10.1051/0004-6361:20021026}, 393, 685

\bibitem[\protect\citeauthoryear{Blasi, Burles  \& Olinto}{Blasi
  et~al.}{1999}]{Blasi1999}
Blasi P.,  Burles S.,   Olinto A.~V.,  1999, \mn@doi [The Astrophysical
  Journal] {10.1086/311958}, 514, L79

\bibitem[\protect\citeauthoryear{Brandenburg \& Subramanian}{Brandenburg \&
  Subramanian}{2005}]{Brandenburg2005}
Brandenburg A.,  Subramanian K.,  2005, \mn@doi [Physics Reports]
  {10.1016/j.physrep.2005.06.005}, 417, 1

\bibitem[\protect\citeauthoryear{Bret}{Bret}{2009}]{Bret2009}
Bret A.,  2009, \mn@doi [The Astrophysical Journal]
  {10.1088/0004-637X/699/2/990}, 699, 990

\bibitem[\protect\citeauthoryear{Broderick, Chang  \& Pfrommer}{Broderick
  et~al.}{2012}]{Broderick2012}
Broderick A.~E.,  Chang P.,   Pfrommer C.,  2012, \mn@doi [The Astrophysical
  Journal] {10.1088/0004-637X/752/1/22}, 752, 22

\bibitem[\protect\citeauthoryear{Brown et~al.,}{Brown et~al.}{2017}]{Brown2017}
Brown S.,  et~al., 2017, \mn@doi [Monthly Notices of the Royal Astronomical
  Society] {10.1093/mnras/stx746}, 468, 4246

\bibitem[\protect\citeauthoryear{Chang, Broderick, Pfrommer, Puchwein,
  Lamberts, Shalaby  \& Vasil}{Chang et~al.}{2016}]{Chang2016}
Chang P.,  Broderick A.~E.,  Pfrommer C.,  Puchwein E.,  Lamberts A.,  Shalaby
  M.,   Vasil G.,  2016, \mn@doi [The Astrophysical Journal]
  {10.3847/1538-4357/833/1/118}, 833, 118

\bibitem[\protect\citeauthoryear{Chluba, Paoletti, Finelli  \&
  Rubi{\~{n}}o-Mart{\'{i}}n}{Chluba et~al.}{2015}]{Chluba2015}
Chluba J.,  Paoletti D.,  Finelli F.,   Rubi{\~{n}}o-Mart{\'{i}}n J.~A.,  2015,
  \mn@doi [Monthly Notices of the Royal Astronomical Society]
  {10.1093/mnras/stv1096}, 451, 2244

\bibitem[\protect\citeauthoryear{Daly \& Loeb}{Daly \& Loeb}{1990}]{Daly1990}
Daly R.~A.,  Loeb A.,  1990, \mn@doi [The Astrophysical Journal]
  {10.1086/169429}, 364, 451

\bibitem[\protect\citeauthoryear{Dermer, Cavadini, Razzaque, Finke, Chiang  \&
  Lott}{Dermer et~al.}{2011}]{Dermer2011}
Dermer C.~D.,  Cavadini M.,  Razzaque S.,  Finke J.~D.,  Chiang J.,   Lott B.,
  2011, \mn@doi [The Astrophysical Journal] {10.1088/2041-8205/733/2/L21}, 733,
  L21

\bibitem[\protect\citeauthoryear{Dolag, Kachelriess, Ostapchenko  \&
  Tom{\`{a}}s}{Dolag et~al.}{2011}]{Dolag2011}
Dolag K.,  Kachelriess M.,  Ostapchenko S.,   Tom{\`{a}}s R.,  2011, \mn@doi
  [The Astrophysical Journal] {10.1088/2041-8205/727/1/L4}, 727, L4

\bibitem[\protect\citeauthoryear{Durrer \& Neronov}{Durrer \&
  Neronov}{2013}]{Durrer2013}
Durrer R.,  Neronov A.,  2013, \mn@doi [The Astronomy and Astrophysics Review]
  {10.1007/s00159-013-0062-7}, 21

\bibitem[\protect\citeauthoryear{Durrive \& Aubert}{Durrive \&
  Aubert}{2017}]{Durrive17}
Durrive J.-B.,  Aubert D.,  2017, in preparation

\bibitem[\protect\citeauthoryear{{Durrive} \& {Langer}}{{Durrive} \&
  {Langer}}{2015}]{Durrive15}
{Durrive} J.-B.,  {Langer} M.,  2015, \mn@doi [\mnras] {10.1093/mnras/stv1578},
  \href {http://adsabs.harvard.edu/abs/2015MNRAS.453..345D} {453, 345}

\bibitem[\protect\citeauthoryear{Evoli \& Ferrara}{Evoli \&
  Ferrara}{2011}]{Evoli2011}
Evoli C.,  Ferrara A.,  2011, \mn@doi [Monthly Notices of the Royal
  Astronomical Society] {10.1111/j.1365-2966.2011.18343.x}, 413, 2721

\bibitem[\protect\citeauthoryear{Falceta-Gon{\c{c}}alves \&
  Kowal}{Falceta-Gon{\c{c}}alves \& Kowal}{2015}]{FalcetaGoncalves15}
Falceta-Gon{\c{c}}alves D.,  Kowal G.,  2015, \mn@doi [The Astrophysical
  Journal] {10.1088/0004-637X/808/1/65}, 808, 65

\bibitem[\protect\citeauthoryear{Fenu, Pitrou  \& Maartens}{Fenu
  et~al.}{2011}]{Fenu2011}
Fenu E.,  Pitrou C.,   Maartens R.,  2011, \mn@doi [Monthly Notices of the
  Royal Astronomical Society] {10.1111/j.1365-2966.2011.18554.x}, 414, 2354

\bibitem[\protect\citeauthoryear{Feretti, Giovannini, Govoni  \&
  Murgia}{Feretti et~al.}{2012}]{Feretti2012}
Feretti L.,  Giovannini G.,  Govoni F.,   Murgia M.,  2012, \mn@doi [The
  Astronomy and Astrophysics Review] {10.1007/s00159-012-0054-z}, 20, 54

\bibitem[\protect\citeauthoryear{Ferrario, Melatos  \& Zrake}{Ferrario
  et~al.}{2015}]{Ferrario2015}
Ferrario L.,  Melatos A.,   Zrake J.,  2015, \mn@doi [Space Science Reviews]
  {10.1007/s11214-015-0138-y}, 191, 77

\bibitem[\protect\citeauthoryear{{Finke}, {Reyes}, {Georganopoulos},
  {Reynolds}, {Ajello}, {Fegan}  \& {McCann}}{{Finke}
  et~al.}{2015}]{2015ApJ...814...20F}
{Finke} J.~D.,  {Reyes} L.~C.,  {Georganopoulos} M.,  {Reynolds} K.,  {Ajello}
  M.,  {Fegan} S.~J.,   {McCann} K.,  2015, \mn@doi [\apj]
  {10.1088/0004-637X/814/1/20}, \href
  {http://adsabs.harvard.edu/abs/2015ApJ...814...20F} {814, 20}

\bibitem[\protect\citeauthoryear{Furlanetto \& Loeb}{Furlanetto \&
  Loeb}{2001}]{Furlanetto2001}
Furlanetto S.~R.,  Loeb A.,  2001, \mn@doi [The Astrophysical Journal]
  {10.1086/321630}, 556, 619

\bibitem[\protect\citeauthoryear{Glover}{Glover}{2013}]{Glover2013}
Glover S.,  2013, in Wiklind T.,  Mobasher B.,   Bromm V.,  eds, Astrophysics
  and Space Science Library, Vol.~396, The First Galaxies -- Theoretical
  Predictions and Observational Clues.
Springer Berlin Heidelberg, pp 103--174

\bibitem[\protect\citeauthoryear{Gluscevic, Venumadhav, Fang, Hirata,
  Oklop{\v{c}}i{\'{c}}  \& Mishra}{Gluscevic et~al.}{2017}]{Gluscevic2017}
Gluscevic V.,  Venumadhav T.,  Fang X.,  Hirata C.,  Oklop{\v{c}}i{\'{c}} A.,
  Mishra A.,  2017, \mn@doi [Physical Review D] {10.1103/PhysRevD.95.083011},
  95, 083011

\bibitem[\protect\citeauthoryear{Gnedin \& Hui}{Gnedin \&
  Hui}{1998}]{Gnedin1998}
Gnedin N.~Y.,  Hui L.,  1998, \mn@doi [Monthly Notices of the Royal
  Astronomical Society] {10.1046/j.1365-8711.1998.01249.x}, 296, 44

\bibitem[\protect\citeauthoryear{{Gnedin}, {Ferrara}  \& {Zweibel}}{{Gnedin}
  et~al.}{2000}]{Gnedin00}
{Gnedin} N.~Y.,  {Ferrara} A.,   {Zweibel} E.~G.,  2000, \mn@doi [\apj]
  {10.1086/309272}, \href {http://adsabs.harvard.edu/abs/2000ApJ...539..505G}
  {539, 505}

\bibitem[\protect\citeauthoryear{Greif, Johnson, Klessen  \& Bromm}{Greif
  et~al.}{2008}]{Greif2008}
Greif T.~H.,  Johnson J.~L.,  Klessen R.~S.,   Bromm V.,  2008, \mn@doi
  [Monthly Notices of the Royal Astronomical Society]
  {10.1111/j.1365-2966.2008.13326.x}, 387, 1021

\bibitem[\protect\citeauthoryear{Gruzinov}{Gruzinov}{2001}]{Gruzinov2001}
Gruzinov A.,  2001, \mn@doi [The Astrophysical Journal] {10.1086/324223}, 563,
  L15

\bibitem[\protect\citeauthoryear{Harrison}{Harrison}{1970}]{Harrison1970}
Harrison E.~R.,  1970, \mn@doi [Monthly Notices of the Royal Astronomical
  Society] {10.1093/mnras/147.3.279}, 147, 279

\bibitem[\protect\citeauthoryear{Harrison}{Harrison}{1973}]{Harrison1973}
Harrison E.~R.,  1973, \mn@doi [Physical Review Letters]
  {10.1103/PhysRevLett.30.188}, 30, 188

\bibitem[\protect\citeauthoryear{{Hayes}, {Schaerer}, {{\"O}stlin},
  {Mas-Hesse}, {Atek}  \& {Kunth}}{{Hayes} et~al.}{2011}]{Hayes11}
{Hayes} M.,  {Schaerer} D.,  {{\"O}stlin} G.,  {Mas-Hesse} J.~M.,  {Atek} H.,
  {Kunth} D.,  2011, \mn@doi [\apj] {10.1088/0004-637X/730/1/8}, \href
  {http://adsabs.harvard.edu/abs/2011ApJ...730....8H} {730, 8}

\bibitem[\protect\citeauthoryear{Iapichino, Schmidt, Niemeyer  \&
  Merklein}{Iapichino et~al.}{2011}]{Iapichino2011}
Iapichino L.,  Schmidt W.,  Niemeyer J.~C.,   Merklein J.,  2011, \mn@doi
  [Monthly Notices of the Royal Astronomical Society]
  {10.1111/j.1365-2966.2011.18550.x}, 414, 2297

\bibitem[\protect\citeauthoryear{{Iwata} et~al.,}{{Iwata}
  et~al.}{2009}]{Iwata09}
{Iwata} I.,  et~al., 2009, \mn@doi [\apj] {10.1088/0004-637X/692/2/1287}, \href
  {http://adsabs.harvard.edu/abs/2009ApJ...692.1287I} {692, 1287}

\bibitem[\protect\citeauthoryear{Kempf, Kilian  \& Spanier}{Kempf
  et~al.}{2016}]{Kempf2016}
Kempf A.,  Kilian P.,   Spanier F.,  2016, \mn@doi [Astronomy {\&}
  Astrophysics] {10.1051/0004-6361/201527521}, 585, A132

\bibitem[\protect\citeauthoryear{{Kim}, {Olinto}  \& {Rosner}}{{Kim}
  et~al.}{1996}]{Kim96}
{Kim} E.-J.,  {Olinto} A.~V.,   {Rosner} R.,  1996, \mn@doi [\apj]
  {10.1086/177667}, \href {http://adsabs.harvard.edu/abs/1996ApJ...468...28K}
  {468, 28}

\bibitem[\protect\citeauthoryear{Kronberg, Lesch  \& Hopp}{Kronberg
  et~al.}{1999}]{Kronberg1999}
Kronberg P.~P.,  Lesch H.,   Hopp U.,  1999, \mn@doi [The Astrophysical
  Journal] {10.1086/306662}, 511, 56

\bibitem[\protect\citeauthoryear{Kulsrud \& Zweibel}{Kulsrud \&
  Zweibel}{2008}]{Kulsrud08}
Kulsrud R.~M.,  Zweibel E.~G.,  2008, \mn@doi [Reports on Progress in Physics]
  {10.1088/0034-4885/71/4/046901}, 71, 046901

\bibitem[\protect\citeauthoryear{Langer, Puget  \& Aghanim}{Langer
  et~al.}{2003}]{Langer03}
Langer M.,  Puget J.-L.,   Aghanim N.,  2003, \mn@doi [Physical Review D]
  {10.1103/PhysRevD.67.043505}, 67, 1

\bibitem[\protect\citeauthoryear{Langer, Aghanim  \& Puget}{Langer
  et~al.}{2005}]{Langer05}
Langer M.,  Aghanim N.,   Puget J.-L.,  2005, \mn@doi [Astronomy {\&}
  Astrophysics] {10.1051/0004-6361:20053372}, 443, 367

\bibitem[\protect\citeauthoryear{Lazar, Schlickeiser, Wielebinski  \&
  Poedts}{Lazar et~al.}{2009}]{Lazar2009}
Lazar M.,  Schlickeiser R.,  Wielebinski R.,   Poedts S.,  2009, \mn@doi [The
  Astrophysical Journal] {10.1088/0004-637X/693/2/1133}, 693, 1133

\bibitem[\protect\citeauthoryear{Loeb \& Furlanetto}{Loeb \&
  Furlanetto}{2013}]{LoebFurlanetto13}
Loeb A.,  Furlanetto S.~R.,  2013, {The First Galaxies in the Universe}.
Princeton Series in Astrophysics, Princeton University Press

\bibitem[\protect\citeauthoryear{McQuinn}{McQuinn}{2016}]{McQuinn2016}
McQuinn M.,  2016, \mn@doi [Annual Review of Astronomy and Astrophysics]
  {10.1146/annurev-astro-082214-122355}, 54, 313

\bibitem[\protect\citeauthoryear{Medvedev, Silva  \& Kamionkowski}{Medvedev
  et~al.}{2006}]{Medvedev2006}
Medvedev M.~V.,  Silva L.~O.,   Kamionkowski M.,  2006, \mn@doi [The
  Astrophysical Journal] {10.1086/504470}, 642, L1

\bibitem[\protect\citeauthoryear{Mellema, Iliev, Pen  \& Shapiro}{Mellema
  et~al.}{2006}]{2006MNRAS.372..679M}
Mellema G.,  Iliev I.~T.,  Pen U.~L.,   Shapiro P.~R.,  2006, \mn@doi [Monthly
  Notices of the Royal Astronomical Society]
  {10.1111/j.1365-2966.2006.10919.x}, 372, 679

\bibitem[\protect\citeauthoryear{Menzler \& Schlickeiser}{Menzler \&
  Schlickeiser}{2015}]{Menzler2015}
Menzler U.,  Schlickeiser R.,  2015, \mn@doi [Monthly Notices of the Royal
  Astronomical Society] {10.1093/mnras/stv232}, 448, 3405

\bibitem[\protect\citeauthoryear{Mishustin \& Ruzmaikin}{Mishustin \&
  Ruzmaikin}{1972}]{Mishustin1972}
Mishustin I.~N.,  Ruzmaikin A.~A.,  1972, Soviet Physics JETP, 34, 233

\bibitem[\protect\citeauthoryear{Mo, van~den Bosch  \& White}{Mo
  et~al.}{2010}]{MoVdBW}
Mo H.,  van~den Bosch F.~C.,   White S. D.~M.,  2010, {Galaxy Formation and
  Evolution}.
Cambridge University Press

\bibitem[\protect\citeauthoryear{Munirov \& Fisch}{Munirov \&
  Fisch}{2017}]{Munirov2017}
Munirov V.~R.,  Fisch N.~J.,  2017, \mn@doi [Physical Review E]
  {10.1103/PhysRevE.95.013205}, 95, 013205

\bibitem[\protect\citeauthoryear{Naoz \& Barkana}{Naoz \&
  Barkana}{2007}]{Naoz2007}
Naoz S.,  Barkana R.,  2007, \mn@doi [Monthly Notices of the Royal Astronomical
  Society] {10.1111/j.1365-2966.2007.11636.x}, 377, 667

\bibitem[\protect\citeauthoryear{Naoz \& Narayan}{Naoz \&
  Narayan}{2013}]{Naoz2013}
Naoz S.,  Narayan R.,  2013, \mn@doi [Physical Review Letters]
  {10.1103/PhysRevLett.111.051303}, 111, 051303

\bibitem[\protect\citeauthoryear{Neronov \& Vovk}{Neronov \&
  Vovk}{2010}]{Neronov2010}
Neronov A.,  Vovk I.,  2010, \mn@doi [Science] {10.1126/science.1184192}, 328,
  73

\bibitem[\protect\citeauthoryear{Oppenheimer \& Dav{\'{e}}}{Oppenheimer \&
  Dav{\'{e}}}{2009}]{Oppenheimer2009}
Oppenheimer B.~D.,  Dav{\'{e}} R.,  2009, \mn@doi [Monthly Notices of the Royal
  Astronomical Society] {10.1111/j.1365-2966.2009.14676.x}, 395, 1875

\bibitem[\protect\citeauthoryear{Pandey \& Sethi}{Pandey \&
  Sethi}{2013}]{Pandey2013}
Pandey K.~L.,  Sethi S.~K.,  2013, \mn@doi [The Astrophysical Journal]
  {10.1088/0004-637X/762/1/15}, 762, 15

\bibitem[\protect\citeauthoryear{{Peebles}}{{Peebles}}{1980}]{Peebles80}
{Peebles} P.~J.~E.,  1980, {The large-scale structure of the universe}.
Princeton University Press

\bibitem[\protect\citeauthoryear{{Planck Collaboration}}{{Planck
  Collaboration}}{2016a}]{Planck15}
{Planck Collaboration} 2016a, \mn@doi [Astronomy {\&} Astrophysics]
  {10.1051/0004-6361/201527101}, 594, A1

\bibitem[\protect\citeauthoryear{{Planck Collaboration}}{{Planck
  Collaboration}}{2016b}]{PlanckCosmoParams2015}
{Planck Collaboration} 2016b, \mn@doi [Astronomy {\&} Astrophysics]
  {10.1051/0004-6361/201525830}, 594, A13

\bibitem[\protect\citeauthoryear{{Planck Collaboration}}{{Planck
  Collaboration}}{2016c}]{PlanckBFields2016}
{Planck Collaboration} 2016c, \mn@doi [Astronomy {\&} Astrophysics]
  {10.1051/0004-6361/201525821}, 594, A19

\bibitem[\protect\citeauthoryear{{Planck Collaboration}}{{Planck
  Collaboration}}{2016d}]{PlanckCollaboration2016}
{Planck Collaboration} 2016d, \mn@doi [Astronomy {\&} Astrophysics]
  {10.1051/0004-6361/201628897}, 596, A108

\bibitem[\protect\citeauthoryear{Press \& Schechter}{Press \&
  Schechter}{1974}]{Press74}
Press W.~H.,  Schechter P.,  1974, \mn@doi [The Astrophysical Journal]
  {10.1086/152650}, 187, 425

\bibitem[\protect\citeauthoryear{Pshirkov, Tinyakov  \& Urban}{Pshirkov
  et~al.}{2016}]{Pshirkov2016}
Pshirkov M.~S.,  Tinyakov P.~G.,   Urban F.~R.,  2016, \mn@doi [Physical Review
  Letters] {10.1103/PhysRevLett.116.191302}, 116, 191302

\bibitem[\protect\citeauthoryear{{Pudritz} \& {Silk}}{{Pudritz} \&
  {Silk}}{1989}]{Pudritz89}
{Pudritz} R.~E.,  {Silk} J.,  1989, \mn@doi [\apj] {10.1086/167625}, \href
  {http://adsabs.harvard.edu/abs/1989ApJ...342..650P} {342, 650}

\bibitem[\protect\citeauthoryear{Raiter, Schaerer  \& Fosbury}{Raiter
  et~al.}{2010}]{Raiter2010}
Raiter A.,  Schaerer D.,   Fosbury R.,  2010, \mn@doi [Astronomy {\&}
  Astrophysics] {10.1051/0004-6361/201015236}, 523, A64

\bibitem[\protect\citeauthoryear{{Rauch}, {Sargent}  \& {Barlow}}{{Rauch}
  et~al.}{2001}]{Rauch2001}
{Rauch} M.,  {Sargent} W.~L.~W.,   {Barlow} T.~A.,  2001, \mn@doi [\apj]
  {10.1086/321402}, \href {http://adsabs.harvard.edu/abs/2001ApJ...554..823R}
  {554, 823}

\bibitem[\protect\citeauthoryear{Ravi et~al.,}{Ravi et~al.}{2016}]{Ravi2016}
Ravi V.,  et~al., 2016, \mn@doi [Science] {10.1126/science.aaf6807}, 354, 1249

\bibitem[\protect\citeauthoryear{Rees}{Rees}{1987}]{Rees1987}
Rees M.~J.,  1987, Royal Astronomical Society, Quarterly Journal, 28, 197

\bibitem[\protect\citeauthoryear{Ryu, Kang  \& Biermann}{Ryu
  et~al.}{1998}]{Ryu1998}
Ryu D.,  Kang H.,   Biermann P.~L.,  1998, Astronomy {\&} Astrophysics, 335, 19

\bibitem[\protect\citeauthoryear{Ryu, Kang, Cho  \& Das}{Ryu
  et~al.}{2008}]{Ryu2008}
Ryu D.,  Kang H.,  Cho J.,   Das S.,  2008, \mn@doi [Science]
  {10.1126/science.1154923}, 320, 909

\bibitem[\protect\citeauthoryear{Ryu, Schleicher, Treumann, Tsagas  \&
  Widrow}{Ryu et~al.}{2012}]{Ryu2012}
Ryu D.,  Schleicher D. R.~G.,  Treumann R.~A.,  Tsagas C.~G.,   Widrow L.~M.,
  2012, \mn@doi [Space Science Reviews] {10.1007/s11214-011-9839-z}, 166, 1

\bibitem[\protect\citeauthoryear{Saga, Ichiki, Takahashi  \& Sugiyama}{Saga
  et~al.}{2015}]{Saga2015}
Saga S.,  Ichiki K.,  Takahashi K.,   Sugiyama N.,  2015, \mn@doi [Physical
  Review D] {10.1103/PhysRevD.91.123510}, 91, 123510

\bibitem[\protect\citeauthoryear{Schaerer}{Schaerer}{2002}]{Schaerer2002}
Schaerer D.,  2002, \mn@doi [Astronomy {\&} Astrophysics]
  {10.1051/0004-6361:20011619}, 382, 28

\bibitem[\protect\citeauthoryear{{Schleicher}, {Banerjee}, {Sur}, {Arshakian},
  {Klessen}, {Beck}  \& {Spaans}}{{Schleicher} et~al.}{2010}]{Schleicher2010}
{Schleicher} D.~R.~G.,  {Banerjee} R.,  {Sur} S.,  {Arshakian} T.~G.,
  {Klessen} R.~S.,  {Beck} R.,   {Spaans} M.,  2010, \mn@doi [\aap]
  {10.1051/0004-6361/201015184}, \href
  {http://adsabs.harvard.edu/abs/2010A%26A...522A.115S} {522, A115}

\bibitem[\protect\citeauthoryear{{Schlickeiser}}{{Schlickeiser}}{2012}]{Schlickeiser2012}
{Schlickeiser} R.,  2012, \mn@doi [Physical Review Letters]
  {10.1103/PhysRevLett.109.261101}, \href
  {http://adsabs.harvard.edu/abs/2012PhRvL.109z1101S} {109, 261101}

\bibitem[\protect\citeauthoryear{{Schlickeiser} \& {Felten}}{{Schlickeiser} \&
  {Felten}}{2013}]{2013ApJ...778...39S}
{Schlickeiser} R.,  {Felten} T.,  2013, \mn@doi [\apj]
  {10.1088/0004-637X/778/1/39}, \href
  {http://adsabs.harvard.edu/abs/2013ApJ...778...39S} {778, 39}

\bibitem[\protect\citeauthoryear{Schlickeiser \& Shukla}{Schlickeiser \&
  Shukla}{2003}]{Schlickeiser2003}
Schlickeiser R.,  Shukla P.~K.,  2003, \mn@doi [The Astrophysical Journal]
  {10.1086/381246}, 599, L57

\bibitem[\protect\citeauthoryear{Schlickeiser \& Yoon}{Schlickeiser \&
  Yoon}{2012}]{Schlickeiser2012a}
Schlickeiser R.,  Yoon P.~H.,  2012, \mn@doi [Physics of Plasmas]
  {10.1063/1.3682985}, 19, 022105

\bibitem[\protect\citeauthoryear{{Schober}, {Schleicher}  \&
  {Klessen}}{{Schober} et~al.}{2013}]{2013A&A...560A..87S}
{Schober} J.,  {Schleicher} D.~R.~G.,   {Klessen} R.~S.,  2013, \mn@doi [\aap]
  {10.1051/0004-6361/201322185}, \href
  {http://adsabs.harvard.edu/abs/2013A%26A...560A..87S} {560, A87}

\bibitem[\protect\citeauthoryear{Sironi \& Giannios}{Sironi \&
  Giannios}{2014}]{Sironi2014}
Sironi L.,  Giannios D.,  2014, \mn@doi [The Astrophysical Journal]
  {10.1088/0004-637X/787/1/49}, 787, 49

\bibitem[\protect\citeauthoryear{{Subramanian}}{{Subramanian}}{2016}]{Subramanian16}
{Subramanian} K.,  2016, \mn@doi [Reports on Progress in Physics]
  {10.1088/0034-4885/79/7/076901}, \href
  {http://adsabs.harvard.edu/abs/2016RPPh...79g6901S} {79, 076901}

\bibitem[\protect\citeauthoryear{Subramanian, Narasimha  \& Chitre}{Subramanian
  et~al.}{1994}]{Subramanian1994}
Subramanian K.,  Narasimha D.,   Chitre S.~M.,  1994, \mn@doi [Monthly Notices
  of the Royal Astronomical Society] {10.1093/mnras/271.1.L15}, 271, L15

\bibitem[\protect\citeauthoryear{Sur, Federrath, Schleicher, Banerjee  \&
  Klessen}{Sur et~al.}{2012}]{Sur2012}
Sur S.,  Federrath C.,  Schleicher D. R.~G.,  Banerjee R.,   Klessen R.~S.,
  2012, \mn@doi [Monthly Notices of the Royal Astronomical Society]
  {10.1111/j.1365-2966.2012.21100.x}, 423, 3148

\bibitem[\protect\citeauthoryear{{Tashiro} \& {Sugiyama}}{{Tashiro} \&
  {Sugiyama}}{2006}]{Tashiro06}
{Tashiro} H.,  {Sugiyama} N.,  2006, \mn@doi [\mnras]
  {10.1111/j.1365-2966.2006.10178.x}, \href
  {http://adsabs.harvard.edu/abs/2006MNRAS.368..965T} {368, 965}

\bibitem[\protect\citeauthoryear{Tavecchio, Ghisellini, Bonnoli  \&
  Foschini}{Tavecchio et~al.}{2011}]{Tavecchio2011}
Tavecchio F.,  Ghisellini G.,  Bonnoli G.,   Foschini L.,  2011, \mn@doi
  [Monthly Notices of the Royal Astronomical Society]
  {10.1111/j.1365-2966.2011.18657.x}, 414, 3566

\bibitem[\protect\citeauthoryear{Tseliakhovich, Barkana  \&
  Hirata}{Tseliakhovich et~al.}{2011}]{Tseliakhovich2011}
Tseliakhovich D.,  Barkana R.,   Hirata C.~M.,  2011, \mn@doi [Monthly Notices
  of the Royal Astronomical Society] {10.1111/j.1365-2966.2011.19541.x}, 418,
  906

\bibitem[\protect\citeauthoryear{Vall{\'{e}}e}{Vall{\'{e}}e}{2011}]{Vallee2011}
Vall{\'{e}}e J.~P.,  2011, \mn@doi [New Astronomy Reviews]
  {10.1016/j.newar.2011.01.002}, 55, 91

\bibitem[\protect\citeauthoryear{Varalakshmi \& Nigam}{Varalakshmi \&
  Nigam}{2017}]{Varalakshmi2017}
Varalakshmi C.,  Nigam R.,  2017, \mn@doi [Astrophysics and Space Science]
  {10.1007/s10509-016-2995-6}, 362, 16

\bibitem[\protect\citeauthoryear{Vazza, Br{\"{u}}ggen, Gheller  \& Wang}{Vazza
  et~al.}{2014}]{Vazza2014}
Vazza F.,  Br{\"{u}}ggen M.,  Gheller C.,   Wang P.,  2014, \mn@doi [Monthly
  Notices of the Royal Astronomical Society] {10.1093/mnras/stu1896}, 445, 3706

\bibitem[\protect\citeauthoryear{Venters \& Pavlidou}{Venters \&
  Pavlidou}{2013}]{Venters2013}
Venters T.~M.,  Pavlidou V.,  2013, \mn@doi [Monthly Notices of the Royal
  Astronomical Society] {10.1093/mnras/stt697}, 432, 3485

\bibitem[\protect\citeauthoryear{Venumadhav, Oklop{\v{c}}i{\'{c}}, Gluscevic,
  Mishra  \& Hirata}{Venumadhav et~al.}{2017}]{Venumadhav2017}
Venumadhav T.,  Oklop{\v{c}}i{\'{c}} A.,  Gluscevic V.,  Mishra A.,   Hirata
  C.~M.,  2017, \mn@doi [Physical Review D] {10.1103/PhysRevD.95.083010}, 95,
  083010

\bibitem[\protect\citeauthoryear{{Wasserman}}{{Wasserman}}{1978}]{Wasserman78}
{Wasserman} I.,  1978, \mn@doi [\apj] {10.1086/156381}, \href
  {http://adsabs.harvard.edu/abs/1978ApJ...224..337W} {224, 337}

\bibitem[\protect\citeauthoryear{Widrow}{Widrow}{2002}]{Widrow2002}
Widrow L.~M.,  2002, \mn@doi [Reviews of Modern Physics]
  {10.1103/RevModPhys.74.775}, 74, 775

\bibitem[\protect\citeauthoryear{Widrow, Ryu, Schleicher, Subramanian, Tsagas
  \& Treumann}{Widrow et~al.}{2012}]{Widrow12}
Widrow L.~M.,  Ryu D.,  Schleicher D. R.~G.,  Subramanian K.,  Tsagas C.~G.,
  Treumann R.~a.,  2012, \mn@doi [Space Science Reviews]
  {10.1007/s11214-011-9833-5}, 166, 37

\bibitem[\protect\citeauthoryear{{Wise} \& {Cen}}{{Wise} \&
  {Cen}}{2009}]{Wise09}
{Wise} J.~H.,  {Cen} R.,  2009, \mn@doi [\apj] {10.1088/0004-637X/693/1/984},
  \href {http://adsabs.harvard.edu/abs/2009ApJ...693..984W} {693, 984}

\bibitem[\protect\citeauthoryear{Wise, Demchenko, Halicek, Norman, Turk, Abel
  \& Smith}{Wise et~al.}{2014}]{Wise14}
Wise J.~H.,  Demchenko V.~G.,  Halicek M.~T.,  Norman M.~L.,  Turk M.~J.,  Abel
  T.,   Smith B.~D.,  2014, \mn@doi [Monthly Notices of the Royal Astronomical
  Society] {10.1093/mnras/stu979}, 442, 2560

\bibitem[\protect\citeauthoryear{Zackrisson, Rydberg, Schaerer, {\"{O}}stlin
  \& Tuli}{Zackrisson et~al.}{2011}]{YggdrasilModel}
Zackrisson E.,  Rydberg C.-E.,  Schaerer D.,  {\"{O}}stlin G.,   Tuli M.,
  2011, \mn@doi [The Astrophysical Journal] {10.1088/0004-637X/740/1/13}, 740,
  13

\bibitem[\protect\citeauthoryear{Zucca, Li  \& Pogosian}{Zucca
  et~al.}{2017}]{Zucca2016}
Zucca A.,  Li Y.,   Pogosian L.,  2017, \mn@doi [Physical Review D]
  {10.1103/PhysRevD.95.063506}, 95, 063506

\makeatother
\end{thebibliography}

\appendix

\section{Evolution of magnetic fields}\label{B-evol}

We provide here the derivation of
equation \eqref{eq:b2pz}, the average magnetic energy density (physical) generated during the EoR.

Assuming magnetic flux freezing, we must account for the fact that any
magnetic field newly generated in the IGM is subsequently adiabatically
diluted by cosmic expansion. Consequently, not taking into account any
possible dissipation or amplification at this point, at a given redshift
$z$, during a redshift interval $dz$ the physical (hence subscripts `p')
magnetic energy density varies as
\begin{equation}
    \dd \left(\frac{B_\mathrm{p}^2}{8\pi}\right) =
     \frac{4}{1+z}\frac{B_\mathrm{p}^2}{8\pi} \dd z + \dd e,
\label{eq:magdiff}
\end{equation}
where the first term on the right-hand side corresponds to adiabatic dilution, and $\dd e$ is a source term, corresponding to the energy density generated during $\dd z$. We have
\begin{equation}
\dd e = \frac{\dd e}{\dd t} \frac{\dd t}{\dd z} \dd z
\label{eq:dE}
\end{equation}
where the time-redshift correspondence is given by the expansion rate at redshift $z$, according to
\begin{equation}
\frac{\dd t}{\dd z} = - \frac{1}{(1+z) H(z)}.
\label{eq:dtoverdz}
\end{equation}
We model the energy generated during $\dd t$ as
\begin{equation}
\frac{\dd {\rm e}}{\dd t} = (1-Q_i)\int^{M_\mathrm{max}}_{M_*} dM E_M ~g_{\rm gl} ~\frac{dn_M}{dM}.
\label{eq:Source}
\end{equation}
Indeed, we construct this term in analogy with the expression for the
volume filling factor \eqref{Qi} but with $E_M$ in place of $V_{\rm
ion}$:  we weigh the number density of haloes by $g_{\rm gl}$ so
that once a source switches on, we add its contribution, but at each
time step, we add the contribution only of the newly born sources as
required. However, equation~\eqref{eq:E_M} for $E_M$ that we
derived in the previous section corresponds to the energy generated in
the IGM by an \textit{isolated} source, while in practice when
considering a distribution of sources, we must take into account the
fact that Str\"omgren spheres overlap. This is essential since our
mechanism is efficient only in neutral regions, so that we expect its
efficiency to decrease as Reionization progresses. Hence, we cannot
simply add up the contribution of sources contained in DM haloes with
equation \eqref{eq:E_M} without care, otherwise we would overestimate the
field generation. We thus introduce in  equation~\eqref{eq:Source} the factor $1-Q_i$, which reduces the fraction of neutral
Hydrogen in the model as time passes, consistently with the amount of
sources switching on since $Q_i$ is given by equation \eqref{Qi}.

Now, the relation given by equation~\eqref{eq:magdiff} yields the differential equation governing the evolution of the physical magnetic energy density. It is convenient to put together the term on the left-hand side with the first term on the right-hand side, and rewrite equation~\eqref{eq:magdiff} as
\begin{equation}
   (1+z)^4 \frac{\dd }{\dd z} \left[(1+z)^{-4} \frac{B_\mathrm{p}^2}{8 \pi}\right]= \frac{\dd  e}{\dd z}
\end{equation}
so that it can be easily integrated. With
equations \eqref{eq:dE}, \eqref{eq:dtoverdz} and \eqref{eq:Source},
we finally get that the mean  magnetic energy density (physical)
generated by photoionizations during EoR evolves with redshift according to equation \eqref{eq:b2pz}.

\bsp

\label{lastpage}

\end{document}